\documentclass[10pt,twocolumn]{IEEEtran}
\usepackage{epsfig,latexsym}
\usepackage{float}
 \usepackage{enumerate}
\usepackage{indentfirst}
\usepackage{amsmath}
\usepackage{amssymb}
\usepackage{times}
\usepackage{subfigure}
\usepackage{cite}
\usepackage{url}
\usepackage{color}
\usepackage{setspace}

\usepackage{amsmath}
\usepackage{amsfonts}
\usepackage{epsfig}
\usepackage{amssymb}
\usepackage{amsmath}
\usepackage{cite}
\usepackage{color,soul}
\usepackage{subfigure}
\usepackage{multirow}
\usepackage{rotating}
\usepackage{graphicx}
\usepackage{tabularx}
\usepackage{array}
\usepackage{color,soul}
\usepackage{graphicx,dblfloatfix}
\usepackage{blindtext}
\usepackage{bigints}
\usepackage{relsize}

\usepackage[hidelinks]{hyperref}
\usepackage{amsthm}
\usepackage{algorithmic}
\usepackage{array}
\usepackage{lipsum}
\usepackage{amsmath}
\usepackage{amsfonts}
\usepackage{epsfig}
\usepackage{amssymb}
\usepackage{graphicx}
\usepackage{amssymb,amsmath}
\usepackage{cite}
\usepackage{color,soul}
\usepackage{amsmath}
\usepackage{color}
\usepackage{multirow}
\usepackage{multicol}
\usepackage{adjustbox}
\usepackage{etoolbox}
\usepackage{amsthm,amssymb}
\usepackage{dblfloatfix}

\begin{document}

\title{Limited Feedback Scheme for \mbox{Device to Device} Communications in 5G Cellular Networks with
Reliability and Cellular Secrecy Outage Constraints}
\author{ Faezeh~Alavi, Nader~Mokari, Mohammad~R.~Javan, and Kanapathippillai~Cumanan} 

\makeatletter
\patchcmd{\@maketitle}
  {\addvspace{0.5\baselineskip}\egroup}
  {\addvspace{-1\baselineskip}\egroup}
  {}
  {}
\makeatother
\maketitle

\begin{abstract}
In this paper, we propose a device to device
(D2D) communication scenario underlaying a cellular
network where both D2D and cellular users (CUs) are discrete
power-rate systems with limited feedback from the receivers.  It is assumed that there exists an adversary which wants to
eavesdrop on the information transmission from the
base station (BS) to CUs.
Since D2D communication shares the same spectrum with cellular
network, cross interference must be considered.
However, when
secrecy capacity is considered, the interference
caused by D2D communication can help to improve the secrecy communications by
confusing the eavesdroppers.
Since both systems share the same spectrum, cross interference must be considered.
We formulate the proposed resource allocation into an optimization problem whose objective is to maximize the average
transmission rate of D2D pair in the presence of the cellular communications under average transmission power constraint.
For the cellular network, we require a minimum average achievable
secrecy rate in the absence of D2D communication as well as a maximum secrecy outage probability in the presence of D2D communication which should be satisfied.
Due to high complexity convex optimization methods, to solve the proposed optimization problem, we apply Particle
Swarm Optimization (PSO) which is an evolutionary approach. Moreover, we model and study the error in the feedback channel and the imperfectness of channel distribution information (CDI) using parametric and nonparametric methods.
Finally, the impact of different system parameters on the performance of the proposed scheme is investigated through simulations.
The performance of the proposed scheme
is evaluated using numerical results for different scenarios.

\end{abstract}

%

\vspace{-0.3cm}
\section{Introduction}\label{introduction}

\vspace{-0.1cm}
\subsection{Background and Motivation}

\vspace{-0.1cm}
The growth of the cellular networks and the number of users as well as the emergence of the new multimedia based services result in growing demands for high data rate and capacities which is beyond the capability of forth generation (4G) wireless networks. Recently, the fifth generation (5G) cellular network has triggered a great attention to provide high data rate and low latency services in a power and spectrally efficient manner. Introducing new applications like context-aware applications requires the direct communications of neighboring devices. In this context, device to device (D2D) communication has been considered as a promising technique for 5G wireless networks \cite{tehrani1,qiao1,yu1}.  
D2D communication operates as an underlay network to a
cellular network \cite{6560489,6750719,6779679} and enables 
reusing the cellular resources which increases the spectral efficiency and the system capacity. In D2D communications, two neighboring devices use the cellular bandwidth to communicate directly without the help of cellular base station (BS).

Although D2D communications can improve the spectral efficiency,
it should provide
access to licensed spectrum with a controlled interference to
avoid the uncertainties of the cellular network performance. Therefore,
interference management is a critical issue for D2D underlaying
cellular networks without considering it, the effectiveness
of D2D communication links will be deteriorated. In
this sense, several papers have proposed mechanisms for
interference mitigation and avoidance. {To perform interference management in D2D underlaying cellular network, one approach is to consider cooperative communications. In this way, a D2D user equipped with multiple antennas acts as an in-band relay to a cellular link where the multi-antenna relay is able to help decoding messages, cancelling interference, and providing multiplexing gain in the network \cite{7368214,7249002,6142093}. }
In \cite{6884156}, power control problem for the D2D users is
investigated in order to optimize the energy efficiency of the
user equipments (UEs) as well as to ensure that
the quality of service (QoS) of D2D
devices and UEs does not fall below the acceptable target. The
problem of interference management through multi rate power
control for D2D communications is
studied in \cite{6666135}. The transmission power levels of D2D users are optimized to
maximize the cell throughput while preserving the
signal-to-interference-plus-noise
ratio (SINR) performance
for the cellular user.
In \cite{6482367}, authors guarantee the reliability of
D2D links and mitigate the interference from the cellular link to
the D2D receivers. A pricing framework has been suggested in
\cite{6884153} where BS protects itself by utilizing game theory approach.

To increase the security of wireless transmission, physical layer security
has been developed based on information theoretic
concepts \cite{7146204,6626661,7047328,7544593,7332326,7784829,7831468,6626307}. From the physical layer point
of view, the security is quantified by the secrecy rate which is defined as the difference of achievable rate between the legitimate receiver and the rate overheard by eavesdroppers \cite{weyner1}.
In this sense, unlike the previous work on D2D underlaying
cellular networks in which the focus is on the interference mitigation and avoidance,
the interference works well when secrecy capacity of the cellular communication is
taken into consideration \cite{6626307}.
In other words, it can be assumed that the D2D communication works as a friendly jammer and its interference is helpful for the secure cellular network to improve secrecy capacity. In practice, since the eavesdropper is a passive attacker, obtaining its
channel state information (CSI) is impossible in many situations. In this case, the secrecy
outage probability can be used as a security performance criterion.

The performance of
previous works is based on the fact that the perfect CSI of all links is available.
However, due to the estimation errors and
feedback delay, perfect CSI may not be available.
In addition, the feedback channel has a limited
capacity since transmitting
unlimited feedback information between transmitters and receivers
means passing a huge amount of bits for signaling.
To tackle this
issue, the limited feedback channel model can be employed. In the limited
feedback channel, the space of channel gains is divided into a finite number of regions, and instead of channel gain values, the index of the
fading region in which the actual channel gain lies is feedbacked \cite{6879482,7100916,6666633}. 
In \cite{6666633}, the authors study the effect of the
feedback information on the performance of the D2D underlaying
cellular networks and develop user selection strategies based on
limited feedback.

\vspace{-0.3cm}
\subsection{Contributions and Organization}

\vspace{-0.1cm}
In this paper, we study D2D communications in the presence of the cellular communications
while there exists a malicious user which wants to eavesdrop the
information transmitted from the BS to CU. We assume that
the legitimate transmitters do not have the perfect values of the channel
power gains and the knowledge about their respective direct
channel power gains is obtained via their dedicated limited rate
feedback channel. In other words, we assume that the space of the channel gains is divided into a finite number of regions. Then given the actual value of the channel gains, the receiver determines the index of the region in which the channel gain lies and feedbacks the index of that region to the corresponding receiver. Note that, the cellular system is superior to D2D communication and D2D pair uses the spectrum of cellular networks in an opportunistic manner. The concurrent transmission of cellular network and D2D pair, if exists, degrades the performance of both systems due to the cross interference between these two systems. Therefore, in this paper, we consider the performance of the cellular system in both the presence and the absence of D2D communication. Precisely, we require that the average transmission rate of cellular user in the absence of the D2D communication should be above a predefined threshold while its performance in the presence of the D2D communication, in terms of outage probability, satisfies a predefined threshold. Our objective is to maximize the average achievable data rate of the D2D pair in the presence of the cellular communication while individual constraints on the average transmission power of the cellular BS and the D2D pair should be satisfied. Due to non-convexity and nonlinearity of the proposed problem, to find the optimal solution of the problem, we use particle swarm optimization (PSO) method which is an evolutionary algorithm \cite{Kennedy,1637688,4358769,494215,7446365}.
In reality, the feedback channels can be affected by the noise which
makes the transmitter select an
incorrect code word from the designed code book. Therefore, in this paper, we consider the effect of error in the feedback channel on the performance of the proposed scheme by incorporating such error into the problem formulation. We further study the effect of channel
distribution information (CDI) imperfectness. Parametric and nonparametric methods are investigated in estimating the CDI of the channels.
The contributions of this paper are as follows:
\begin{itemize}
\item
We develop a mathematical model for the secure communication in D2D communication underlaying the cellular network in which the knowledge of transmitters about the CSIs is obtained via a limited rate feedback channel. In our model, we consider the cross interference between the cellular network and the D2D pair explicitly and formulate the resource allocation problem as an optimization framework.

\item To solve this optimization problem and obtain its solutions which are the fading regions' boundaries and transmission power levels, we use PSO algorithm which is an evolutionary algorithm.

\item We further consider the effect of the noise in the feedback channel and incorporate it into our optimization problem. In this case, the error in the feedback channel would lead the transmitters to choose the incorrect code-words. We formulate the corresponding optimization problem and solve it using the PSO approach.

\item We also consider the effect of the CDI imperfectness in our proposed scheme. In this case, the CDI's parameters are not perfectly known and parametric and non-parametric approaches are used to estimate the CDI parameters.
\end{itemize}
Finally, the performance of the proposed scheme in different scenarios is investigated via simulations.

The paper is organized as follows. System model
is described in Section \ref{sysmodel}. Limited rate feedback schemes are proposed in Section \ref{limitedratefeedbackschemes}. The limited rate feedback resource
allocation problem is formulated and solved in Section \ref{resourceallocationproblem}. In Section \ref{practicalconsideration},
practical considerations, i.e., noisy feedback channel and CDI estimation error, are investigated. Simulation and numerical results are provided in
Section \ref{simulationresults} and finally conclusions are drawn in Section \ref{conclusions}.

\vspace{-0.3cm}
\section{System Model}\label{sysmodel}

\vspace{-0.2cm}
We consider a D2D communication scenario underlaying an existing
cellular network. It is assumed the downlink transmission in the cellular
network where the BS transmits
information to a cellular user while at the same time; the
existing D2D pair performs its own transmission on the same
channel. Such a scenario can be interpreted as there are many
cellular users in the network each of which is assigned to a
channel over which the BS sends information to them. Assuming this
assignment is performed based on some network parameters and is
fixed, two cellular users exploit one of the available cellular
channels to perform their information transmission directly. In this paper, we assume that this assignment is predefined.
In addition to cellular user and D2D pair, we assume that there
exists a malicious user which wants to eavesdrop on the information
transmission of cellular network, i.e., from the BS to the CU. However, the malicious user does not eavesdrop on the D2D pair.
Such assumption can be justified when the malicious user is not
aware of the existence of D2D pair as such sharing can be
performed opportunistically (i.e., D2D pair may or may not
exist at any time) when the malicious user is not interested in
D2D pair information, or when the D2D pair applies upper layer
security measures, e.g., cryptography. In such case, the malicious
user treats the signals from D2D pair as noise.

\begin{figure}[t]
  \begin{center}
    \includegraphics[scale=0.35]{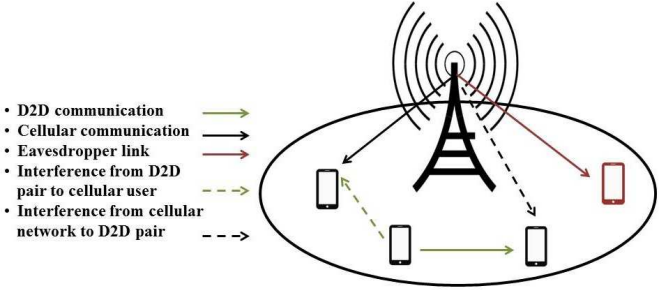} 
    \caption{A D2D communication underlaying an existing cellular network. }
 \vspace{-0.5cm}
      \label{systemmodel}
  \end{center}
\end{figure}

Let $h^{\text{BC}}$, $h^{\text{BD}}$, $h^{\text{DD}}$,
$h^{\text{DC}}$, $h^{\text{BE}}$, and $h^{\text{DE}}$ denote,
respectively, the noise normalized channel power gain of the
channel from BS to CU, from BS to the receiver of D2D
pair (RD2D), from the transmitter of D2D pair (TD2D) to RD2D, from TD2D to CU, from
BS to the eavesdropper, and from the TD2D to the
eavesdropper. We assume that all channels undergo independent
block fading with Rayleigh distribution meaning that the channel
power gains, i.e., $h^{\text{BC}}$, $h^{\text{BD}}$,
$h^{\text{DD}}$, $h^{\text{DC}}$, $h^{\text{BE}}$, and
$h^{\text{DE}}$, are exponentially distributed with the mean of
$\bar{h}^{\text{BC}}$, $\bar{h}^{\text{BD}}$,
$\bar{h}^{\text{DD}}$, $\bar{h}^{\text{DC}}$,
$\bar{h}^{\text{BE}}$, and
$\bar{h}^{\text{DE}}$, respectively.

In this paper, we assume that the
eavesdropper has complete knowledge about the instantaneous channel power gains and
the CDI of the channels from BS to
CU and from BS to itself. We further assume
that the legitimate receivers, i.e., the cellular user and the
RD2D, know the CDI of all channels and only the
instantaneous channel power gains of their respective channels.
We assume that, the legitimate transmitters do not have perfect
values of channel power gains and the knowledge of legitimate
transmitters about their respective direct channel power gains is
obtained via their respective limited rate feedback channels. In
this case, the space of $h^{\text{BC}}$ is divided into
a finite number of $M$ regions, i.e., $[0,\widetilde{h}^{\text{BC}}(1))$,
$[\widetilde{h}^{\text{BC}}(1),\widetilde{h}^{\text{BC}}(2))$,
$\cdots$, $[\widetilde{h}^{\text{BC}}(M-1),
\widetilde{h}^{\text{BC}}(M))$ where
$\widetilde{h}^{\text{BC}}(M)=\infty$. Similarly, for D2D pair,
the space of $h^{\text{DD}}$ is divided into a finite number of $N$ regions, i.e., $[0,\widetilde{h}^{\text{DD}}(1))$,
$[\widetilde{h}^{\text{DD}}(1),\widetilde{h}^{\text{DD}}(2))$,
$\cdots$, $[\widetilde{h}^{\text{DD}}(N-1),
\widetilde{h}^{\text{DD}}(N))$ where
$\widetilde{h}^{\text{DD}}(N)=\infty$. The receiver, i.e.,
CU, measures the channel power gain $h^{\text{BC}}$ and
feedbacks the index $m$ if $h^{\text{BC}}$ lies in the region
$[\widetilde{h}^{\text{BC}}(m),\widetilde{h}^{\text{BC}}(m+1))$.
Similarly, RD2D measures the channel power
gain $h^{\text{DD}}$ and feedbacks the index $n$ if $h^{\text{DD}}$
lies in the region
$[\widetilde{h}^{\text{DD}}(n),\widetilde{h}^{\text{DD}}(n+1))$.
In this paper, we assume that the feedback links are
confidential. This means that, the feedbacked index of the cellular
network could not be overheard by D2D pair and that of D2D pair
could not be overheard by the cellular network. In this case, the
power-rate tuples will depend only on the corresponding feedbacked
index, i.e., we have $(p^{\text{BC}}(m),
r^{\text{BC}}(m),r_\text{S}^{\text{BC}}(m))$ for cellular
communication which are the transmit power, the transmission rate and the secrecy
rate at which BS transmits information to CU, respectively.
Moreover, we consider $(p^{\text{DD}}(n), r^{\text{DD}}(n))$ for D2D communications
which are the transmit power and the transmission rate at which TD2D transmits information to RD2D, respectively.

{In fact, the proposed schemes operate in two phases.  In the first phase (off-line phase), several parameters that are later used for resource allocation are computed. It is done before the communication established and based on the CDIs of the network's links the optimum boundary regions and code-books are designed by the base station. At the end of this phase, all code-words are informed to users by BS. However, the channel partitioning structure is kept at the CU. In the second phase (on-line phase) that is employed during communication, the transmitters use the parameters obtained in off-line phase. In fact, in the on-line phase, the CU and D2D receiver measure the related CSIs and based on them find the related channel partition and boundary region.  Then, they transmit back the index to the base station and D2D transmitter in order to select the corresponding code-word from the obtained code-book. Note that the code-book, which contains a set of code-words, is designed off-line and known by each node. The computational burden takes place during the initialization (off-line) phase and requires a negligible burden during the transmission (online) phase and it is certainly desirable from an implementation perspective \cite{giannakis1,7347459,7446365}. In this paper, we consider that there is a central processing unit and some assignments are performed in this step. Then, the network uses the information prepared in the central processing  \cite{6805125,5208020,5073611}.}

{For the resource allocation, two approaches could be adopted. One is to consider the problem of pairing the D2D and cellular links as well as the designing limited feedback scheme jointly. In this way, the outcome of the resource allocation problem is which D2D link is paired with which cellular link as well as code-books (power allocation and boundary regions). However, this approach is much complex and would be computationally prohibitive. Another approach which could lower the complexity of the scheme is to consider the pairing problem and limited rate feedback design separately. In this case, one first solves the problem of pairing D2D and cellular link. Then, given this pairing result, the problem of designing a limited rate feedback scheme could be formulated and solved. In this paper, we assumed the second approach and assumed that the D2D and cellular links are paired and the pairing result is available based on which we design the limited rate feedback scheme. The pairing process could be performed based on network parameters as well some degrees of the required QoS level. For example, one could formulate a problem in which the aim is to pair the D2D and cellular links based on the average channel gains instead of instantaneous or long term channel considerations.  Several authors have studied the problem of pairing D2D links with CUs for spectrum sharing and focus on the selection of the D2D link and CUs as a pair for better performance \cite{7437034,6926861,6731768}.  However, in our paper, we present the resource allocation in D2D underlaying cellular network and focus on devising a limited rate feedback model as well as power allocation problem encompassing different performance metrics. In this way, first, the D2D link and cellular communication link are scheduled based on the mean of channel power gain. In the next step, the resource allocation can be obtained based on the proposed scheme in this paper. To obtain the best optimum solution, they should be solved at the same time; however, it causes a high computational complexity. To reduce the complexity, they can be considered separately at the cost of a slight performance loss. However, we can extend this approach for solving the problem at the same time as future works.}

\section{Limited Rate Feedback Schemes}\label{limitedratefeedbackschemes}
\subsection{Capacity of Links}\label{capacityoflinks}
When the feedback links are confidential, the feedbacked indices
cannot be heard by any party other than the eavesdropper. In this
case, we assume that, the cellular network quantizes the main
channel power gain, i.e., $h^{\text{BC}}$, independent of the
index feedbacked by the RD2D.
Knowing that the index
$m$ is feedbacked by CU, from the designed code
book $\mathcal{C}^\text{BC}$, BS chooses transmit power level $p^{\text{BC}}(m)$ to
send its information. In other word, it chooses
the tuple
$(p^{\text{BC}}(m), r^{\text{BC}}(m),r_\text{S}^{\text{BC}}(m))$
where $r^{\text{BC}}(m)=\log(1+\widetilde{h}^{\text{BC}}(m)
p^{\text{BC}}(m))$ is the transmission rate over BS to CU.
In this case, in
the absence of D2D transmission, the capacity of the link between
BS and CU and its corresponding secrecy capacity
are, respectively, given by

\vspace{-0.5cm}
\begin{eqnarray}\label{capacitycellulardirectlinkd2dabsence}
C^{\text{C}}(m)=\log(1+h^{\text{BC}}
p^{\text{BC}}(m)),
\end{eqnarray}

\vspace{-0.6cm}
\begin{align}\label{secrecycapacitycellulard2dabsence}
C_\text{S}^{\text{C}}(m)=\big[\log(1+h^{\text{BC}}
p^{\text{BC}}(m))-\log(1+h^{\text{BE}} p^{\text{BC}}(m))\big]^+\!\!\!,
\end{align}

\vspace{-0.2cm}
\noindent while knowing indices n, the D2D pair chooses transmit power level $p^{\text{DD}}(n)$
to send its information. In other word, the D2D pair
has chosen $(p^{\text{DD}}(n),
r^{\text{DD}}(n))$ for concurrent transmission with cellular
network and the capacity of the link between BS and CU
and its corresponding secrecy capacity are, respectively, given by

\vspace{-0.5cm}
\begin{eqnarray}\label{capacitycellulardirectlinkd2dpresence}
\hat{C}^{\text{C}}(m,n)=\log(1+\hat{h}^{\text{BC}}
p^{\text{BC}}(m)),
\end{eqnarray}

\vspace{-0.6cm}
\begin{align}\label{secrecycapacitycellulard2dpresence}
\hat{C}_\text{S}^{\text{C}}(m,n)=\big[\log(1+\hat{h}^{\text{BC}}
p^{\text{BC}}(m))-\log(1+\hat{h}^{\text{BE}}
p^{\text{BC}}(m))\big]^+\!\!\!,
\end{align}

\vspace{-0.4cm}
\noindent where
$\hat{h}^{\text{BC}}=\frac{h^{\text{BC}}}{1+h^{\text{DC}}p^{\text{DD}}(n)}$
and
$\hat{h}^{\text{BE}}=\frac{h^{\text{BE}}}{1+h^{\text{DE}}p^{\text{DD}}(n)}$
are the effective channel gains between BS and
CU and between BS and eavesdropper, respectively. Note that, the transmission capacity in \eqref{capacitycellulardirectlinkd2dpresence} and the secrecy capacity in \eqref{secrecycapacitycellulard2dpresence} can be achieved only if we have full knowledge of CSIs, i.e., the perfect values of $\hat{h}^{\text{BC}}$ and $\hat{h}^{\text{BE}}$.

On the other hand, given that the index $n$ is feedbacked by RD2D, from the designed code book
$\mathcal{C}^\text{DD}$, TD2D chooses the
tuple $(p^{\text{DD}}(n), r^{\text{DD}}(n))$ where
$r^{\text{DD}}(n)=\log(1+\widetilde{h}^{\text{DD}}(n)
p^{\text{DD}}(n))$. In this case, in the absence of cellular
transmission, the capacity of D2D link is given by
\begin{eqnarray}\label{capacityD2Ddirectlinkcellularabsence}
C^{\text{D}}(n)=\log(1+h^{\text{DD}} p^{\text{DD}}(n)),
\end{eqnarray}
while given that BS has chosen the tuple $(p^{\text{BC}}(m),
r^{\text{BC}}(m),r_\text{S}^{\text{BC}}(m))$ for concurrent
transmission with D2D pair, the capacity of the D2D link is given
by
\begin{eqnarray}\label{capacityD2Ddirectlinkcellularpresence}
\hat{C}^{\text{D}}(m,n)=\log(1+\hat{h}^{\text{DD}}
p^{\text{DD}}(n)),
\end{eqnarray}
where
$\hat{h}^{\text{DD}}=\frac{h^{\text{DD}}}{1+h^{\text{BD}}p^{\text{BC}}(m)}$
is the effective channel gain between TD2D and RD2D. The perfect value of $\hat{h}^{\text{DD}}$ is needed to achieve the transmission capacity in \eqref{capacityD2Ddirectlinkcellularpresence}.

\vspace{-0.5cm}
\subsection{Outage Events}\label{outageevent}

\vspace{-0.1cm}
In our model, there are two types of outage, namely reliability
outage which corresponds to the case where the transmission rate
exceeds the channel capacity and secrecy outage whose definition depends on the availability of CSI at the transmitter. More precisely, consider the case where CU feedbacks the index $m$, i.e., BS chooses the tuple
$(p^{\text{BC}}(m), r^{\text{BC}}(m),r_\text{S}^{\text{BC}}(m))$.
In the absence of D2D transmission, reliability outage for
cellular communication occurs if $r^{\text{BC}}(m) >
C^{\text{C}}(m)$ where $C^{\text{C}}(m)$ is given by
(\ref{capacitycellulardirectlinkd2dabsence}). This event
corresponds to the case where $\widetilde{h}^{\text{BC}}(m) > h^{\text{BC}}$
which never occurs. In addition, the secrecy outage occurs if
$r_\text{S}^{\text{BC}}(m)
> C_\text{S}^{\text{C}}(m)$ where $C_\text{S}^{\text{C}}(m)$ is
given by (\ref{secrecycapacitycellulard2dabsence}). In this paper,
however, we are interested in the outage event in the presence of
D2D pair communication. In this case, reliability outage for
cellular communication occurs if $r^{\text{BC}}(m) >
\hat{C}^{\text{C}}(m,n)$ where $\hat{C}^{\text{C}}(m,n)$ is given
by (\ref{capacitycellulardirectlinkd2dpresence}). This event
corresponds to the case where $\widetilde{h}^{\text{BC}}(m) >
\hat{h}^{\text{BC}}$ which is possible. However, as we assume that only the knowledge of direct channels is available, the secrecy
outage does not correspond to the event $r_\text{S}^{\text{BC}}(m)
> \hat{C}_\text{S}^{\text{C}}(m,n)$ where $\hat{C}_\text{S}^{\text{C}}(m,n)$ is
given by (\ref{secrecycapacitycellulard2dpresence}). Note that, given that the indices $m$ and $n$ are feedbacked, the transmission rate is fixed to $r^{\text{BC}}(m)=\log(1+\widetilde{h}^{\text{BC}}(m)
p^{\text{BC}}(m))$. In this case, any secrecy rate given by $r_\text{S}^{\text{BC}}(m)\leq \Big(r^{\text{BC}}(m) -r^{\text{e}}(m)\Big)$ is achievable where $r^{\text{e}}(m)$ is the maximum allowable equivocation rate of the eavesdropper. Now, assume for the feedbacked index $m$, the secrecy rate is fixed to $r_\text{S}^{\text{BC}}(m)$ and hence we have $r^{\text{e}}(m)=\Big(r^{\text{BC}}(m) -r_\text{S}^{\text{BC}}(m)\Big)$. Therefore, the secrecy outage occurs if the instantaneous capacity of the eavesdropper exceeds the value of $r^{\text{e}}(m)$, i.e., we have $\hat{C}^\text{BE}(m,n)=\log(1+\hat{h}^{\text{BE}}
p^{\text{BC}}(m))>r^{\text{e}}(m)$ where the dependence of the value of $\hat{C}^\text{BE}(m,n)$ on the feedbacked index $n$ is through $\hat{h}^{\text{BE}}=\frac{h^{\text{BE}}}{1+h^{\text{DE}}p^{\text{DD}}(n)}$.     Therefore,
given that D2D pair chooses $(p^{\text{DD}}(n),
r^{\text{DD}}(n))$, the outage probability for cellular
communication using tuple $(p^{\text{BC}}(m),
r^{\text{BC}}(m),r_\text{S}^{\text{BC}}(m))$ is given by
\begin{align}\nonumber
&P^\text{outage}_{p^{\text{BC}}(m),
r^{\text{BC}}(m),r_\text{S}^{\text{BC}}(m),p^{\text{DD}}(n),
r^{\text{DD}}(n)}=\\\label{outageprobabilitycellularregionmn}
&\qquad\qquad\qquad 1-P^\text{success}_{p^{\text{BC}}(m),
r^{\text{BC}}(m),r_\text{S}^{\text{BC}}(m),p^{\text{DD}}(n),
r^{\text{DD}}(n)},
\end{align}

\vspace{-0.4cm}
\noindent where
\begin{align}\nonumber
&P^\text{success}_{p^{\text{BC}}(m),
r^{\text{BC}}(m),r_\text{S}^{\text{BC}}(m),p^{\text{DD}}(n),
r^{\text{DD}}(n)}=\\\label{successprobabilitycellularregionmn}
&\Pr\bigg(r^{\text{BC}}(m)\leq
\hat{C}^{\text{C}}(m,n),\hat{C}^\text{BE}(m,n)\leq
r^{\text{BC}}(m)-r_\text{S}^{\text{BC}}(m)\bigg),
\end{align}

\vspace{-0.3cm}
\noindent and we assumed that $h^{\text{BC}}\in
[\widetilde{h}^{\text{BC}}(m),\widetilde{h}^{\text{BC}}(m+1))$ and $h^{\text{DD}}\in
[\widetilde{h}^{\text{DD}}(n),\widetilde{h}^{\text{DD}}(n+1))$.

\vspace{0.1cm}
Using the above explanations and defining $\mathcal{R}^{\text{BC}}_m= [\widetilde{h}^{\text{BC}}_m,
\widetilde{h}^{\text{BC}}_{m+1})$ and $\mathcal{R}^{\text{DD}}_n= [\widetilde{h}^{\text{DD}}_n,
\widetilde{h}^{\text{DD}}_{n+1})$, the outage probability for cellular
communication when it uses the tuple $(p^{\text{BC}}(m),
r^{\text{BC}}(m),r_\text{S}^{\text{BC}}(m))$ is given by

\vspace{-0.5cm}
\begin{align}\nonumber
&P^\text{outage}_{p^{\text{BC}}(m),
r^{\text{BC}}(m),r_\text{S}^{\text{BC}}(m)}=\\\label{outageprobabilitycellularregionm}
&\sum_{n=1}^{N-1}\Pr\bigg(h^{\text{DD}}\in
\mathcal{R}^{\text{DD}}_n\bigg)P^\text{outage}_{p^{\text{BC}}(m),
r^{\text{BC}}(m),r_\text{S}^{\text{BC}}(m),p^{\text{DD}}(n),
r^{\text{DD}}(n)},
\end{align}

\vspace{-0.3cm}
\noindent and the outage probability of cellular link code book, i.e.,
$\mathcal{C}^\text{BC}$, is given by
\begin{eqnarray}\label{outageprobabilitycellularcodebook}
P^\text{outage}_{\mathcal{C}^\text{BC}}=\sum_{m=1}^{M-1}\Pr\bigg(h^{\text{BC}}\in\mathcal{R}^{\text{BC}}_m\bigg)P^\text{outage}_{{p^\text{BC}}(m),
r^{\text{BC}}(m),r_\text{S}^\text{BC}(m)}.
\end{eqnarray}

\vspace{-0.2cm}
Details on obtaining the above probabilities are deferred to Appendix \ref{findingprobabilityoutageprobabilitycellularregionmn}.

Similarly, for the D2D pair, assume that the transmitter chooses
the pair $(p^{\text{DD}}(n), r^{\text{DD}}(n))$. In the absence of
cellular transmission, reliability outage occurs if
$r^{\text{DD}}(n) > C^{\text{D}}(n)$ where $C^{\text{D}}(m)$ is
given by (\ref{capacityD2Ddirectlinkcellularabsence}). This event
corresponds to the case where $\widetilde{h}^{\text{DD}}(n) >
h^{\text{DD}}$ which never occurs. On the other hand, in the
presence of cellular communication, reliability outage for D2D
communication occurs if $r^{\text{DD}}(n) >
\hat{C}^{\text{D}}(m,n)$ where $\hat{C}^{\text{D}}(m,n)$ is given
by (\ref{capacityD2Ddirectlinkcellularpresence}). This event
corresponds to the case where $\widetilde{h}^{\text{DD}}(n) >
\hat{h}^{\text{DD}}$ which is possible. Note that, as we assumed
 the malicious user is not interested in D2D communication,
only reliable transmission is considered for D2D pair and no
secrecy rate is defined.

\vspace{-0.4cm}
\subsection{Transmit Powers and Achievable Rates}\label{transmitpowersandachievablerates}

\vspace{-0.1cm}
As we assumed, the transmitters only know the region number in
which the channel power gains of direct channels lay. This means
that it is impossible to know the value of effective channel gains
when a concurrent transmission is running. Therefore, for cellular
network the value of direct channel gain, i.e., $h^{\text{BC}}$,
and for D2D pair, the value of $h^{\text{DD}}$ are quantized.
Given that $h^{\text{BC}}$ lies in the region
$\mathcal{R}^{\text{BC}}_m$,
BS chooses tuple $(p^{\text{BC}}(m),
r^{\text{BC}}(m),r_\text{S}^{\text{BC}}(m))$. Note that,
regardless of the channel power gains of other links, i.e.,
$h^{\text{BD}}$, $h^{\text{DD}}$, $h^{\text{DC}}$,
$h^{\text{BE}}$, and $h^{\text{DE}}$, BS transmits with power
level $p^{\text{BC}}(m)$. Therefore, in this case, the average
transmission power of BS only depends on $h^{\text{BC}}$ and is
given by

\vspace{-0.5cm}
\begin{eqnarray}\label{averagepowercellular}
\bar{P}^{\text{C}}=\sum_{m=1}^{M-1}\Pr\bigg(h^{\text{BC}}\in \mathcal{R}^{\text{BC}}_m \bigg)p^{\text{BC}}(m),
\end{eqnarray}

\vspace{-0.2cm}
\noindent which is the same for both cases where D2D pair is not
transmitting or concurrently transmits information. Similarly,
given that $h^{\text{DD}}$ lies in the region
$\mathcal{R}^{\text{DD}}_n$, TD2D chooses the pair $(p^{\text{DD}}(n),
r^{\text{DD}}(n))$. As we assumed that the feedback link of
cellular network is confidential, the transmission power of D2D pair,
i.e., $p^{\text{DD}}(n)$, only depends on $h^{\text{DD}}$.
Therefore, in this case, the average transmission power of D2D
pair is given by

\vspace{-0.5cm}
\begin{eqnarray}\label{averagepowerD2Dconficetial}
\bar{P}^{\text{D}}=\sum_{n=1}^{N-1}\Pr\bigg(h^{\text{DD}}\in \mathcal{R}^{\text{DD}}_n \bigg)p^{\text{DD}}(n),
\end{eqnarray}

\vspace{-0.2cm}
\noindent which does not depend on whether BS is transmitting concurrently
or not.

In addition, for cellular communication, the transmission is assumed successful if no outage occurs, i.e., we
have both the reliable and secure communications. We define the
average achievable secrecy rate for cellular communication as the
adopted secrecy rate, i.e., $r_\text{S}^{\text{BC}}(m)$, times the
probability of success, i.e., no outage occurs, summed over all
regions. When the D2D pair is absent, the average achievable
secrecy rate is given by

\vspace{-0.5cm}
\begin{eqnarray}\label{averageratecellulard2dabsent}
\bar{R}_\text{S}^{\text{C}}=\sum_{m=1}^{M-1}\Pr\bigg(h^{\text{BC}}\in \mathcal{R}^{\text{BC}}_m ,r_\text{S}^{\text{BC}}(m)\leq
C_\text{S}^{\text{C}}(m)\bigg)r_\text{S}^{\text{BC}}(m),
\end{eqnarray}

\vspace{-0.2cm}
\noindent where $C_\text{S}^{\text{C}}(m)$ is given by
(\ref{secrecycapacitycellulard2dabsence}). Note that, it is not
required to include the term $r^{\text{BC}}(m)\leq
C^{\text{C}}(m)$ in (\ref{averageratecellulard2dabsent}) because
it is always satisfied. Please refer to Appendix \ref{findingsuccessprobabilityinaverageratecellulard2dabsent} for more details on obtaining the probability terms in \eqref{averageratecellulard2dabsent}.

Since, we need only reliable
transmission for D2D communication, the average achievable rate
of D2D pair is defined as the adopted data rate. i.e.,
$r^{\text{DD}}(n)$, times the probability of succeed, i.e., no
outage occurs, summed over all region.
The average transmission rate which is achievable by D2D pair in
the presence of cellular communication is given by
\begin{align}\nonumber
&\bar{R}^{\text{D}}= \sum_{m=1}^{M-1}\sum_{n=1}^{N-1} \\\label{averagerateD2Dcellularpresentconficetial}
&\Pr\bigg(h^{\text{BC}}\in
\mathcal{R}^{\text{BC}}_m ,h^{\text{DD}}\in
\mathcal{R}^{\text{DD}}_n ,r^{\text{DD}}(n)\leq
\hat{C}^{\text{D}}(m,n)\bigg)r^{\text{DD}}(n),
\end{align}

\vspace{-0.4cm}
\noindent where $\hat{C}^{\text{D}}(m,n)$ is given by
(\ref{capacityD2Ddirectlinkcellularpresence}). Details on obtaining probability terms in \eqref{averagerateD2Dcellularpresentconficetial} can be found in Appendix \ref{findingprobabilityaveragerateD2Dcellularpresentconficetial}.

\vspace{-0.3cm}
\section{Limited Rate Feedback Resource Allocation Problem}\label{resourceallocationproblem}
%
\subsection{Problem Formulation}\label{problemformulation}

\vspace{-0.1cm}
In this paper, we assume that D2D pair opportunistically uses the
cellular network resources to maximize its average transmission
rate. Note that, generally, D2D communication is underlay to
cellular communication which means the later one is superior
and should be protected against the side effects of concurrent
transmission of D2D pair.  There are several approaches to achieve
this. One approach is to limit the amount of interference that D2D pair
produces on the cellular receiver. Such approach can be seen exactly
the same as the notion of interference temperature in cognitive
radio networks \cite{6884153}. However, note that this approach is effective when
it is used in its instantaneous form (i.e., the exact amount of
interference D2D pair produces) and not the averaged one (i.e.,
the average amount of interference D2D pair produces). However,
since we only know the direct channel power gains using limited
rate feedback, applying instantaneous interference constraint is
not possible. Another approach is to maintain the average
achievable rate of the cellular link above a predefined threshold \cite{6666135,6884156}.
Moreover, the reliability of the cellular network is much of our concern, particularly,
in the case that the resource is shared with D2D links. To this end, outage based approach is the next
approach in which the outage probability for cellular
communication is kept below a predefined threshold \cite{6666135,6482367}.

In this paper, we combine the last two approaches. More precisely,
our objective is to maximize the average achievable data rate for D2D pair in the presence of
cellular communication, i.e.,
\eqref{averagerateD2Dcellularpresentconficetial}, while it is required
to maintain a minimum amount of the average achievable data rate
of cellular link in the absence of  D2D communication, i.e.,
\eqref{averageratecellulard2dabsent}, and the outage probability
for cellular communication in
the presence of D2D
communication, i.e., \eqref{outageprobabilitycellularcodebook},
is kept below a predefined threshold. In this way, we take into
account the performance of cellular communication both in the absence
and presence of D2D communication. Indeed, by doing so, we
require that the average transmission rate of cellular link in the
absence of D2D pair to stay above a predefined threshold while its
performance in the presence of
D2D communication, which is given by the outage probability, remains as satisfactory as is required.  In addition, the average transmit power of the cellular link and D2D pair, which are, respectively, given by \eqref{averagepowercellular} and
\eqref{averagepowerD2Dconficetial}, should not exceed a predefined value. Mathematically, defining $\mathcal{A}=\bigg\{\widetilde{\textbf{h}}^{\text{BC}},\widetilde{\textbf{h}}^{\text{DD}},\textbf{p}^{\text{BC}},
\textbf{p}^{\text{DD}},\textbf{r}_\text{S}^{\text{BC}}\bigg\}$, we aim to solve the optimization problem which is given by

\vspace{-0.4cm}
\begin{subequations} \label{mainprob}
\begin{align}\nonumber
&\max_{\mathcal{A}}
\sum_{m=1}^{M-1}\sum_{n=1}^{N-1}\Pr\bigg(h^{\text{BC}}\in
\mathcal{R}^{\text{BC}}_m,
h^{\text{DD}}\in\mathcal{R}^{\text{DD}}_n \\\label{optimizationproblem1}
&\quad\quad\qquad\quad\qquad\qquad ,r^{\text{DD}}(n)\leq
\hat{C}^{\text{D}}(m,n)\bigg)r^{\text{DD}}(n),\\
\label{averageratecellular1}
&\text{s.t.:} \sum_{m=1}^{M-1}\!\Pr\!\bigg(h^{\text{BC}}\!\in
\mathcal{R}^ {\text{BC}}_m ,r_\text{S}^{\text{BC}}(m)\leq
C_\text{S}^{\text{C}}(m)\!\bigg)r_\text{S}^{\text{BC}}(m)\!\geq\!
\bar{R}_\text{S}^{\text{Cmin}}\!\!\!\!,\\
\label{outageprobabilitycellular1}
&\sum_{m=1}^{M-1}\Pr\big(
h^{\text{BC}}\!\in \!\mathcal{R}^{\text{BC}}_m
\big)P^\text{outage}_{p^{\text{BC}}(m),
r^{\text{BC}}(m),r_\text{S}^{\text{BC}}(m)}\!\!\leq
\!P^\text{outage,max}_{\mathcal{C}^\text{BC}}\!\!\!,\\ \label{averagepowercellular1}
&\sum_{m=1}^{M-1}\Pr\big(h^{\text{BC}}\in
\mathcal{R}^{\text{BC}}_m \big)p^{\text{BC}}(m)\leq
\bar{P}^{\text{C,max}},\\\label{averagepowerd2d1}
&\sum_{n=1}^{N-1}\Pr\big(h^{\text{DD}}\in
\mathcal{R}^{\text{DD}}_n \big)p^{\text{DD}}(n)\leq
\bar{P}^{\text{D,max}}.
\end{align}
\end{subequations}

This optimization problem is nonlinear
and non-convex and it is hard to solve it, hence, we utilize the PSO method
which has been used to solve
highly non-linear mixed integer optimization problems in various research \cite{1637688,4358769,494215,7446365}.
Since PSO is a computational intelligence-based technique and has global search ability, it can converge to the optimal solution and not largely affected by the size and non-linearity of the problem \cite{4358769}.

\vspace{-0.4cm}
\subsection{Particle Swarm Optimization Method}\label{PSO}
In this paper, to solve the optimization problem, we apply relatively new technique,
PSO algorithm which
is a computational intelligence-based technique.
PSO is based on a moment of the swarm which searches to find the best optimal solution by updating generations \cite{Kennedy,7446365}. This method is
not largely affected by the size and nonlinearity of the problem,
and can converge to the optimal solution.
In PSO algorithm, all particles which are the potential solutions, move towards its optimum value. For each
iteration all the particles in this swarm are updated by its position and velocity for
optimization ability and based on them the aim function for the system is evaluated.
PSO starts with the random initialization of swarm of particles in the search space. Then,
by adjusting the path of each particle to its own best location
and the best particle of the swarm at each step, the global best
solution is found. The path of each particle in the search space
is adjusted by its velocity, according to moving experience of
that particle and other particles in the search space.

In this paper, we consider different particles for each variable, i.e, $\mathcal{A}=\bigg\{\widetilde{\textbf{h}}^{\text{BC}},\widetilde{\textbf{h}}^{\text{DD}},\textbf{p}^{\text{BC}},
\textbf{p}^{\text{DD}},\textbf{r}_\text{S}^{\text{BC}}\bigg\}$, which denote a solution of the problem.
The PSO algorithm consists of $\mathcal{A}_i$ as the
vector of $i^{th}$ particle in $d$ dimension, i.e, for $\{\widetilde{\textbf{h}}^{\text{BC}},
\textbf{p}^{\text{BC}}, \textbf{r}_\text{S}^{\text{BC}}\}$, $d$ is equal to $M-1$ and for  $\{\widetilde{\textbf{h}}^{\text{DD}}, \textbf{p}^{\text{DD}}\}$, $d$ is equal to $N-1$ \cite{7446365}.

The position and the velocity of the $i^{th}$ particle in the $d$ dimensional
search space can be shown as $X_i=[x_{i,1}, x_{i,2},\ldots , x_{i,d}]^T$
and $V_i=[v_{i,1}, v_{i,2},\ldots , v_{i,d}]^T$, respectively.
A best position of each particle is denoted by pbest \big($P_i=[p_{i,1}, p_{i,2},\ldots , p_{i,d}]^T $\big), corresponding to the personal best objective
value obtained at time $t$. The global best particle, i.e., gbest ($p_g$), shows the best particle at time $t$ in the entire swarm. The new velocity
of each particle can be obtained as follows \cite{7446365}:
\begin{align}\nonumber
 &v_{i,j}(t+1)=wv_{i,j}(t)+c_1r_1(p_{i,j}-x_{i,j}(t))\\\label{pso}
 &+c_2r_2(p_{g}-x_{i,j}(t)),\quad j=1,\ldots,d,
\end{align}

\vspace{-0.2cm}
\noindent where $c_1$ and $c_2$ are constants called acceleration coefficients,
$w$ is the inertia factor, $r_1$ and $r_2$ are two
independent random numbers uniformly distributed in $[0, 1]$.
Thus, the position of each particle is updated in each
step as follows:
\begin{equation}\label{update}
  x_{i,j}(t+1)=x_{i,j}(t)+v_{i,j}(t+1).
\end{equation}

\vspace{-0.2cm}
The standard form of PSO uses \eqref{pso} to calculate the new
velocity of each particle based on its previous velocity and the
distance of its current position from both its best position and global best position.
To control search of
particles outside the search space $[X_{i}^{min}\!\!,X_{i}^{max}] $, we can limit the value of $V_i$ to the range $[V_{i}^{min}\!\!,V_{i}^{max}] $ and according to \eqref{update}, each
particle moves to a new position.
The process is repeated until a stopping
criterion is satisfied. This algorithm is summarized in Table.\ref{PSOscheme} \cite{Kennedy}.

\begin{table}[t!]
\caption{PSO scheme for our problem }\label{PSOscheme}
\vspace{-0.5cm}
\begin{center}
\begin{tabular}{|p{8cm}|}
  \hline
  \textbf{Initialization:}~ \\
  \textbf{Step 1)} MaxIt: Iteration number of PSO algorithm \\
  \qquad \quad \;\, nPop: Number of particles of PSO algorithm \\
  \qquad \quad \;\, For each variable of $\mathcal{A}=\bigg\{\widetilde{\textbf{h}}^{\text{BC}},\widetilde{\textbf{h}}^{\text{DD}},\textbf{p}^{\text{BC}},
\textbf{p}^{\text{DD}},\textbf{r}_\text{S}^{\text{BC}}
\bigg\}$:\\
  \qquad \qquad \quad $X_i$: Position of one particle,  $i= 1,2,\ldots ,\text{nPop}$\\
  \qquad \qquad \quad $V_i$: Velocity of one particle, $i= 1,2,\ldots ,\text{nPop}$\\
   \textbf{Step 2)} Evaluate \eqref{optimizationproblem1}-\eqref{averagepowerd2d1} as a cost for all particles, named $cost_{i}$:\\
   \qquad \quad \;\, Set $pbest_i=X_i$ and $pbest.cost_i=cost_i$ \\
   \qquad \quad \;\, Set $gbest$ and $gbest.cost$ value equal to the value of the best \\
   \qquad \quad \;\, initial particle. \\ \hline
   For $t= 1,2,\ldots ,\text{MaxIt}$\\
   \quad For $i= 1,2,\ldots ,\text{nPop}$\\
   \qquad \qquad  Use \eqref{pso}, \eqref{update} to update the velocity and position of particles \\
   \qquad \qquad  for all variables of $\mathcal{A}=\bigg\{\widetilde{\textbf{h}}^{\text{BC}},\widetilde{\textbf{h}}^{\text{DD}},\textbf{p}^{\text{BC}},
\textbf{p}^{\text{DD}},\textbf{r}_\text{S}^{\text{BC}}\bigg\}$ \\
  \qquad \qquad Evaluate \eqref{optimizationproblem1}-\eqref{averagepowerd2d1} \\
  \qquad \qquad If $cost_{i} > {pbest.cost}_{i}$ : \\
   \qquad \qquad \; $pbest_i=X_i$ and ${pbest.cost}_{i}=cost_i$.\\
    \qquad \quad \; \; If ${pbest.cost}_{i} > gbest.cost$ \\
     \qquad \quad \; \; \; $gbest=pbest_i$ and $gbest.cost=pbest.cost_i$.\\
  \quad end \\
  end \\
  \hline
\end{tabular}
\end{center}
\end{table}

\vspace{-0.4cm}
\section{Practical Consideration}\label{practicalconsideration}

\vspace{-0.1cm}
\subsection{CDI Estimation Error}\label{cdiestimatinerror}
The most practical assumption made in this paper is that the
instantaneous channel power gains of the eavesdropper's links, i.e.,
$h^{\text{BE}}$ and $h^{\text{DE}}$, are not available which is
mostly due to the fact that the eavesdropper is passive and hence
acquiring its channel power gains are not possible. Generally, the
CDI of a channel depends on the environmental property of the communication channel. If the propagation environment is known,
one can assume that the channel CDIs, including those of the eavesdropper, are available. The statistical property of the
signal propagation in the coverage area of the network can be
easily obtained as the legitimate users are present and can be
involved in finding the required statistical properties. Since for
small geographical areas, a unified distribution can be applied to
all channels\footnote{This assumption is reasonable when the size
of the area under investigation is small which is the case for
nowadays cellular networks specially for small cells.}, we can have the CDIs of eavesdropper's links
at hand.

Due to the availability of limited statistical data, the
distribution function is hard to drive and cannot be fit into the known ones, e.g., Rayleigh distribution. In such cases, schemes
developed based on the availability of the perfect CDI may exhibit
performance worse than that expected. Therefore, the imperfectness
of CDIs should be taken into account. Generally, such
consideration can be performed by assuming that the true
distribution differs from the nominal distribution by the value
known as Kullback--Leibler distance \cite{cover1} and incorporate
such inaccuracy into problem formulation
\cite{ioannou1,gong1,li1}. We investigate imperfect CDI through two parametric and nonparametric methods.

\subsubsection{Parametric Method}
In parametric methods, the effect of the imperfect CDI is studied through the performance loss by simulations as in Section \ref{simulationresults}. This means that,
we solve the optimization problem (\ref{optimizationproblem1})
with the available channel CDIs and obtain the channel
quantization and code books for cellular link and D2D pair. Then,
we evaluate the performance loss due to imperfect CDI in terms of
changes in the average achievable rates. In other words, we consider
the imperfect channel power gain of each channel $i$ which
is exponentially distributed with the mean of $\ddot{\bar{h}}^{\text{i}}$ :

\vspace{-0.4cm}
\begin{equation}\label{param}
 \ddot{\bar{h}}^{\text{i}}=(1-\Delta) \bar{h}^{\text{i}},
\end{equation}

\vspace{-0.1cm}
\noindent where $\Delta$ is percent error of imperfect CDI.


\subsubsection{Non-Parametric Method}
Another way to estimate CDI is nonparametric method which estimates the density based on the
received samples from the channel.
In this paper, we adopt two nonparametric methods: kernel density estimation (KDE) and robust KDE (RKDE).

\textit{2.1. Kernel Density Estimation
(KDE):}
One of the most well-known non-parametric density
estimation methods is kernel density estimation \cite{Scott2010}.
When the samples, referred to as the nominal data, are noise free, KDE can provide a good estimate
of the density. A set of observations $\{x_1,...,x_L\}\in \mathbb{R}^j$ is used to
estimate a random vector $x$  with a density $f(x)$ where $L$ is
the number of observation vectors. Moreover, each $x_i = {x_{i1},
. . . , x_{ij}}, i = 1, . . . ,L$ is a sequence of $j$ data in the
vector $x_i$. The kernel density estimate of $f(x)$ given by

\vspace{-0.3cm}
\begin{equation}\label{kde}
\hat{f}_{KDE}(x)=\frac{1}{L}\sum_{i=1}^L k_\delta(x, x_i),
\end{equation}

\vspace{-0.2cm}
\noindent where $k_\delta(x, x_i)$ is the kernel function which commonly is a Gaussian kernel:

\vspace{-0.5cm}
\begin{equation}\label{kernel}
 k_\delta(x, x_i)=(\frac{1}{\sqrt{2\Pi}\delta})^j exp(-\frac{\parallel
 x-x_i\parallel^2}{2\delta^2}),
\end{equation}

\vspace{-0.1cm}
\noindent where $\delta$ is the smoothing parameter and referred to as the bandwidth.
It is set to the median distance
of a training point $x_i$ to its nearest neighbor.

\textit{2.2. Robust Kernel Density
Estimation:} In practice, the
channel gain samples might include contaminated data, referred to as outlier data,
which makes it necessary to use robust density estimation methods such as robust KDE (RKDE).
In the presence of the contaminated samples, RKDE can give
robustness to contamination of the training sequence and estimate
the density. Contaminated data consists of realizations from both a
nominal or clean distribution in addition to outlying or anomalous
measurements. In an increasing number of applications, data arises
from high dimensional or high-throughput systems where the nominal
distribution itself may be quite complex and not amenable
to parametric modelling. 
The RKDE has the following form:

\vspace{-0.3cm}
\begin{equation}\label{RKDE}
\hat{f}_{RKDE}(x)=\sum_{i=1}^L \omega_i k_\delta(x, x_i),
\end{equation}

\vspace{-0.2cm}
\noindent where $k_\delta(x, x_i)$ is a kernel function and $\omega_i$ are nonnegative
weights that sum to one.
The RKDE can be implemented based on the iteratively reweighed least
square (IRWLS) \cite{6952401} algorithm in which the main goal is to find
the optimal value of $\omega_i$.

\vspace{-0.2cm}
\subsection{Noisy Feedback Channel}\label{noisyfeedbackchannel}

\vspace{-0.1cm}
So far, we assumed that the feedback channels are error free
meaning that the received index is the same as the feedbacked one.
However, in reality, the feedback channel could be affected by
the noise which makes transmitter to select an incorrect code word from
the designed code book. Note that, designing limited rate feedback
systems with incorporating feedback error is complicated,
especially for our scheme with two interfering links. In this
paper, to consider the feedback error, we utilize the
scheme which is commonly used in the literature
\cite{ekbatani1,he3,nader2,6678798,7446365}.
We consider the memoryless feedback channel which
characterized by index transition probabilities
$\rho^\text{C}_{m,m'}$ ($m,m'=0,\cdots,M-1$) for cellular link
which is the probability of receiving index $m$ in BS given the
index $m'$ was sent by CU, and $\rho^\text{C}_{n,n'}$
($n,n'=0,\cdots,N-1$) for D2D pair which is the probability of
receiving index $n$ in TD2D given the index
$n'$ was sent by RD2D. It is assumed
$b_M=\log_2(M)$ bits feedback for cellular link and $b_N=\log_2(N)$
bits feedback for D2D pair. Let $m_1 m_2 \cdots m_{b_M}$, $m'_1
m'_2 \cdots m'_{b_M}$, $n_1 n_2 \cdots n_{b_M}$, and $n'_1 n'_2
\cdots n'_{b_M}$ indicate the binary
display of indices $m$, $m'$, $n$, and $n'$, respectively. We assume that
the cellular and D2D pair's feedback channel can be considered as,
respectively, $b_M$ and $b_N$ independent use
of binary symmetric channel (BSC) to sent each of the feedback
bits presented in binary representations of cellular link and D2D
pair's feedbacked indices. Let $q^\text{C}$ and $q^\text{D}$
represent the cross over probabilities of the feedback channels of
cellular link and D2D pair, respectively. The index
transition probabilities of the feedback channels of cellular link
and D2D pair can be obtained, respectively, by

\vspace{-0.5cm}
\begin{eqnarray}\label{indextransitionprobabilitycellular}
\rho^\text{C}_{m,m'}=(q^\text{C})^{d_{m,m'}}(1-q^\text{C})^{b_M-d_{m,m'}},\\\label{indextransitionprobabilitycellular}
\rho^\text{D}_{n,n'}=(q^\text{D})^{d_{n,n'}}(1-q^\text{D})^{b_N-d_{m,m'}},
\end{eqnarray}

\vspace{-0.2cm}
\noindent where $d_{m,m'}$ and $d_{n,n'}$ denote the Hamming distances
between, indices $m$ and $m'$ and indices $n$ and
$n'$, respectively \cite{ekbatani1,he3,nader2}.

With the above definitions and assumptions, the average transmission
powers in (\ref{averagepowercellular}) and
(\ref{averagepowerD2Dconficetial}), average transmission data
rates in (\ref{averageratecellulard2dabsent}) and
(\ref{averagerateD2Dcellularpresentconficetial}), and the outage
probabilities in (\ref{outageprobabilitycellularregionmn}),
(\ref{successprobabilitycellularregionmn}),
(\ref{outageprobabilitycellularregionm}), and
(\ref{outageprobabilitycellularcodebook}) should be manipulated to
incorporate the effect of noisy feedback channel. Note that,
choosing the transmit power level from a code book only depends on
the corresponding channel region index which is feedbacked by the
respective transmitter. This mean that, in
(\ref{averagepowercellular}), we should only consider the noise
effect of the feedback channel of cellular link, and in
(\ref{averagepowerD2Dconficetial}), we should only consider
the noise effect of the feedback channel of D2D pair. Therefore,
the average transmit powers of cellular link and D2D pair, when
noisy channel feedback is assumed, are given, respectively, by
\vspace{-0.3cm}
\begin{align}\label{averagepowercellularnoisyfeedback}
&\bar{P}^{\text{C}}=\sum_{m=1}^{M-1}\sum_{m'=1}^{M-1}\rho^\text{C}_{m,m'}\Pr\bigg(h^{\text{BC}}\in \mathcal{R}^{\text{BC}}_{m'} \bigg)p^{\text{BC}}(m),\\
\label{averagepowerD2Dconficetialnoisyfeedback}
&\bar{P}^{\text{D}}=\sum_{n=1}^{N-1}\sum_{n'=1}^{N-1}\rho^\text{D}_{n,n'}
\Pr\bigg(h^{\text{DD}}\in \mathcal{R}^{\text{DD}}_{n'} \bigg)p^{\text{DD}}(n).
\end{align}

\vspace{-0.2cm}
For the average transmission data rate in
(\ref{averageratecellulard2dabsent}), we
assumed the D2D pair is absent, hence, it is not affected by the noise in feedback channel of D2D pair.
Therefore, the average transmission data rate can be written as

\vspace{-0.4cm}
\begin{align}\nonumber
&\bar{R}_\text{S}^{\text{C}}=\!\sum_{m=1}^{M-1}\!\sum_{m'=1}^{M-1}
\!\Pr\!\bigg(h^{\text{BC}}\!\in \!\mathcal{R}^{\text{BC}}_{m'},
m'\!\rightarrow \!m,r^{\text{BC}}(m)\!\leq\!
C^{\text{C}}(m/m') \\\label{averageratecellulard2dabsentnoisyfeedback}
&\qquad\qquad\qquad\qquad,{C}^\text{BE}(m)\leq
r^{\text{e}}(m)\bigg)r_\text{S}^{\text{BC}}(m),
\end{align}

\vspace{-0.3cm}
\noindent where $ r^{\text{e}}(m)=r^{\text{BC}}(m)-r_\text{S}^{\text{BC}}(m)$ and $C^{\text{BE}}(m)=\log(1+{h}^{\text{BE}}
p^{\text{BC}}(m))$ where we note that actually, the value of $C^{\text{BE}}(m)$ does not depend on $m'$. In \eqref{averageratecellulard2dabsentnoisyfeedback}, $ m'\rightarrow m$ is the event that the feedbacked index $m'$ is received as $m$. Here, we highlight that, in
(\ref{averageratecellulard2dabsentnoisyfeedback}), $C^{\text{C}}(m/m')$ means that its value is given by \eqref{capacitycellulardirectlinkd2dabsence} with $h^{\text{BC}}\in \mathcal{R}^{\text{BC}}_{m'}$
and $p^{\text{BC}}(m)$. Note that, here, in contrast to \eqref{averageratecellulard2dabsent}, we must include $r^{\text{BC}}(m)\leq
C^{\text{C}}(m/m')$ in (\ref{averageratecellulard2dabsentnoisyfeedback}) because reliability outage can occur when the feedback is noisy.

From \eqref{capacitycellulardirectlinkd2dabsence}, we know that the event $r^{\text{BC}}(m)\leq
C^{\text{C}}(m/m')$ occurs when $m\leq m'$. Therefore, \eqref{averageratecellulard2dabsentnoisyfeedback} can be rewritten as follows:

\vspace{-0.5cm}
\begin{align}\label{averageratecellulard2dabsentnoisyfeedbackriwritten}
&\bar{R}_\text{S}^{\text{C}}\!=\!\!\!\sum_{m=1}^{M-1}\!\!\sum_{m'=m}^{M-1}
\!\!\rho^\text{C}_{m,m'}\!\Pr\!\bigg(\!h^{\text{BC}}\!\in\!
\mathcal{R}^{\text{BC}}_{m'}
,\hat{C}^\text{BE}(m)\!\leq\!
r^{\text{e}}(m)\!\bigg)\!r_\text{S}^{\text{BC}}(m).
\end{align}

\vspace{-0.2cm}
The remaining steps are similar to those in obtaining probability terms in \eqref{averageratecellulard2dabsent} in Appendix \ref{findingsuccessprobabilityinaverageratecellulard2dabsent}, and hence omitted.
However, as we assumed in
(\ref{averagerateD2Dcellularpresentconficetial}) that both the
cellular link and D2D pair transmit simultaneously, the cross
effect of noisy feedback channel should be considered. In other
words, given that the transmitted index $m'$ was received as $m$ by
BS and the transmitted index $n'$ was received as $n$ by
transmitter of D2D pair, the average data rate of D2D pair in the presence of cellular
communication with noisy
feedback channels is given by

\vspace{-0.5cm}
\begin{align}\nonumber
&\bar{R}^{\text{D}}\!=\!\sum_{m=1}^{M-1}\!\sum_{m'=1}^{M-1}\!\sum_{n=1}^{N-1}\!\sum_{n'=1}^{N-1} \Pr\big(h^{\text{BC}}\in\mathcal{R}^{\text{BC}}_{m'},h^{\text{DD}}\in\mathcal{R}^{\text{DD}}_{n'}, \\\label{averagerateD2Dcellularpresentconficetialnoisyfeedback}
&\hspace{0.4cm}(m',n')\!\rightarrow\!(m,n),r^{\text{DD}}(n)\!\leq
\!\hat{C}^{\text{D}}(m,n/n')\!\big)r^{\text{DD}}(n).
\end{align}

\vspace{-0.1cm}
Note that, in
(\ref{averagerateD2Dcellularpresentconficetialnoisyfeedback}), the value of $\hat{C}^{\text{D}}(m,n/n')$ does not depend on $m'$. In addition, the
effect of noise in feedback channel of cellular link on the value
of $\hat{C}^{\text{D}}(m,n/n')$ appears through the choice of
transmit power level $p^{\text{BC}}(m)$ which affects the value of
the effective channel gain
$\hat{h}^{\text{DD}}=\frac{h^{\text{DD}}}{1+h^{\text{BD}}p^{\text{BC}}(m)}$
with $h^{\text{DD}}\in\mathcal{R}^{\text{DD}}_{n'}$. Obtaining probability terms in \eqref{averagerateD2Dcellularpresentconficetialnoisyfeedback} is similar to obtaining probability terms in \eqref{averagerateD2Dcellularpresentconficetial} in Appendix \ref{findingprobabilityaveragerateD2Dcellularpresentconficetial}, and hence omitted.

Like
(\ref{averagerateD2Dcellularpresentconficetialnoisyfeedback}), for
the outage probabilities in
(\ref{outageprobabilitycellularregionmn}),
(\ref{successprobabilitycellularregionmn}),
(\ref{outageprobabilitycellularregionm}), and
(\ref{outageprobabilitycellularcodebook}), we should consider the
cross effect of noisy feedback channels. If we consider the noisy
feedback channel effect in the outage probability of cellular
communication in the presence of D2D pair, we observe that given
the feedback indices $m'$ and $n'$ were received by the
corresponding receiver as $m$ and $n$, respectively, BS uses the
code word $(p^{\text{BC}}(m),
r^{\text{BC}}(m),r_\text{S}^{\text{BC}}(m))$ while we have
$h^{\text{BC}}\in\mathcal{R}^{\text{BC}}_{m'}$ and the transmitter
of D2D pair uses the code word $p^{\text{DD}}(n),
r^{\text{DD}}(n)$ while we have
$h^{\text{DD}}\in\mathcal{R}^{\text{DD}}_{n'}$. In this case, the
outage probability is given by

\vspace{-0.5cm}
\begin{align}\nonumber
&P^{\text{outage}(m/m',n)}_{p^{\text{BC}}(m),
r^{\text{BC}}(m),r_\text{S}^{\text{BC}}(m),p^{\text{DD}}(n),
r^{\text{DD}}(n)}=\\\label{outageprobabilitycellularregionmnnoisyfeedback}
&1-P^{\text{success}(m/m',n)}_{p^{\text{BC}}(m),
r^{\text{BC}}(m),r_\text{S}^{\text{BC}}(m),p^{\text{DD}}(n),
r^{\text{DD}}(n)},
\end{align}

\vspace{-0.3cm}
\noindent where

\vspace{-0.3cm}
\begin{align}\nonumber
&P^{\text{success}(m/m',n)}_{p^{\text{BC}}(m),
r^{\text{BC}}(m),r_\text{S}^{\text{BC}}(m),p^{\text{DD}}(n),
r^{\text{DD}}(n)}\!=\\\label{successprobabilitycellularregionmnnoisyfeedback}
&\!\Pr\!\bigg(\!r^{\text{BC}}(m)\!\leq\!
\hat{C}^{\text{C}}(m/m',n),\hat{C}^\text{BE}(m,n)\leq
r^{\text{BC}}(m)\!-\!r_\text{S}^{\text{BC}}(m)\!\bigg),
\end{align}

\vspace{-0.2cm}
\noindent where $\hat{C}^{\text{C}}(m/m',n)$ is given by
(\ref{capacitycellulardirectlinkd2dpresence}) with
$\hat{h}^{\text{BC}}=\frac{h^{\text{BC}}}{1+h^{\text{DC}}p^{\text{DD}}(n)}$ and $h^{\text{BC}}\in \mathcal{R}^{\text{BC}}_{m'}$, and $\hat{C}^\text{BE}(m,n)=\log\big(1+\hat{h}^{\text{BE}}
p^{\text{BC}}(m)\big)$ with
$\hat{h}^{\text{BE}}=\frac{h^{\text{BE}}}{1+h^{\text{DE}}p^{\text{DD}}(n)}$. To obtain \eqref{successprobabilitycellularregionmnnoisyfeedback} one can follow the similar steps as those for \eqref{successprobabilitycellularregionmn} in Appendix \ref{findingprobabilityoutageprobabilitycellularregionmn}.

Using the above explanations, the outage probability for cellular
communication when it uses the tuple
$(p^{\text{BC}}(m),r^{\text{BC}}(m),r_\text{S}^{\text{BC}}(m))$
and under noisy feedback channel model, is given by

\vspace{-0.5cm}
\begin{align}\nonumber
P^{\text{outage}(m/m')}_{p^{\text{BC}}(m),
r^{\text{BC}}(m),r_\text{S}^{\text{BC}}(m)}=&\sum_{n=1}^{N-1}\sum_{n'=1}^{N-1}\Big(\rho^\text{D}_{n,n'} \Pr\big(h^{\text{DD}}\in \mathcal{R}^{\text{DD}}_{n'} \big)\\\label{outageprobabilitycellularregionmnoisyfeedback}
&\!\!\!\!\!\!\!\!\!\!\!\!\!\!\!\!\!\!\!\times P^{\text{outage}(m/m',n)}_{p^{\text{BC}}(m),
r^{\text{BC}}(m),r_\text{S}^{\text{BC}}(m),p^{\text{DD}}(n),
r^{\text{DD}}(n)}\Big),
\end{align}

\vspace{-0.1cm}
\noindent and the outage probability of cellular link code book, i.e.,
$\mathcal{C}^\text{BC}$, is given by

\vspace{-0.3cm}
\begin{align}\label{outageprobabilitycellularcodebooknoisyfeedback}
&P^\text{outage}_{\mathcal{C}^\text{BC}}\!=\!\!\sum_{m=1}^{M-1}\!\sum_{m'=1}^{M-1}\!
\rho^\text{C}_{m,m'}\!\Pr\!\big(h^{\text{BC}}\!\in \! \mathcal{R}^{\text{BC}}_{m'}\!\big)~P^\text{outage}_{p^{\text{BC}}(m),
r^{\text{BC}}(m),r_\text{S}^{\text{BC}}(m)}.
\end{align}

\vspace{-0.2cm}
To take the noisy feedback channel model into consideration, in
the optimization problem (\ref{optimizationproblem1}), we must use
(\ref{averagepowercellularnoisyfeedback}),
(\ref{averagepowerD2Dconficetialnoisyfeedback}),
(\ref{averageratecellulard2dabsentnoisyfeedback}),
(\ref{averagerateD2Dcellularpresentconficetialnoisyfeedback}), and
(\ref{outageprobabilitycellularcodebooknoisyfeedback}).

\vspace{-0.2cm}
\section{Simulation Results}\label{simulationresults}
In this section, numerical results are presented to evaluate the
performance of the proposed limited feedback scheme in a D2D communication
through the simulations under various system parameters. The channel gain is an
exponential random variable with the probability density function
(PDF) given by

\vspace{-0.2cm}
\begin{equation}
  f(h)=\frac{1}{\sigma}\exp(\frac{-h}{\sigma}),
\end{equation}

\vspace{-0.1cm}
\noindent where $\sigma$
can be used to model the average channel
gain as $\sigma=s(\frac{d}{d_0})^{-\gamma}$ where $d$ is the distance between the transmitter and the receiver, $d_0$ is the reference distance, $\gamma$ is
the amplitude path-loss exponent, and $s$ characterizes the
shadowing effect.
The users are assumed to be uniformly
distributed in a cell of radius $100$ m. The small-scale channel fading is
assumed to be Rayleigh distributed.
The path-loss exponent is equal to $4$, and
the shadowing effect follows a log-normal distribution, i.e.,
$10\log_{10}(s)\sim N(0,8 dB) $.
System parameters are equal to $P_\text{D}^{\text{max}}=10$ dB, $P_\text{C}^{\text{max}}=5$ dB, $P_{\text{outage}}^{\text{max}}=0.1,$ ${R^\text{C}_\text{S}}_{\text{min}}=0.1$ bps/Hz, $q^C=q^D=0.25$.
We set the coefficients $c_1=c_2=1.496$ and $w=0.729$ for PSO algorithm
and simulated for $1000$ iterations.

\begin{figure}[t!]
  \begin{center}
    \includegraphics[scale=0.2]{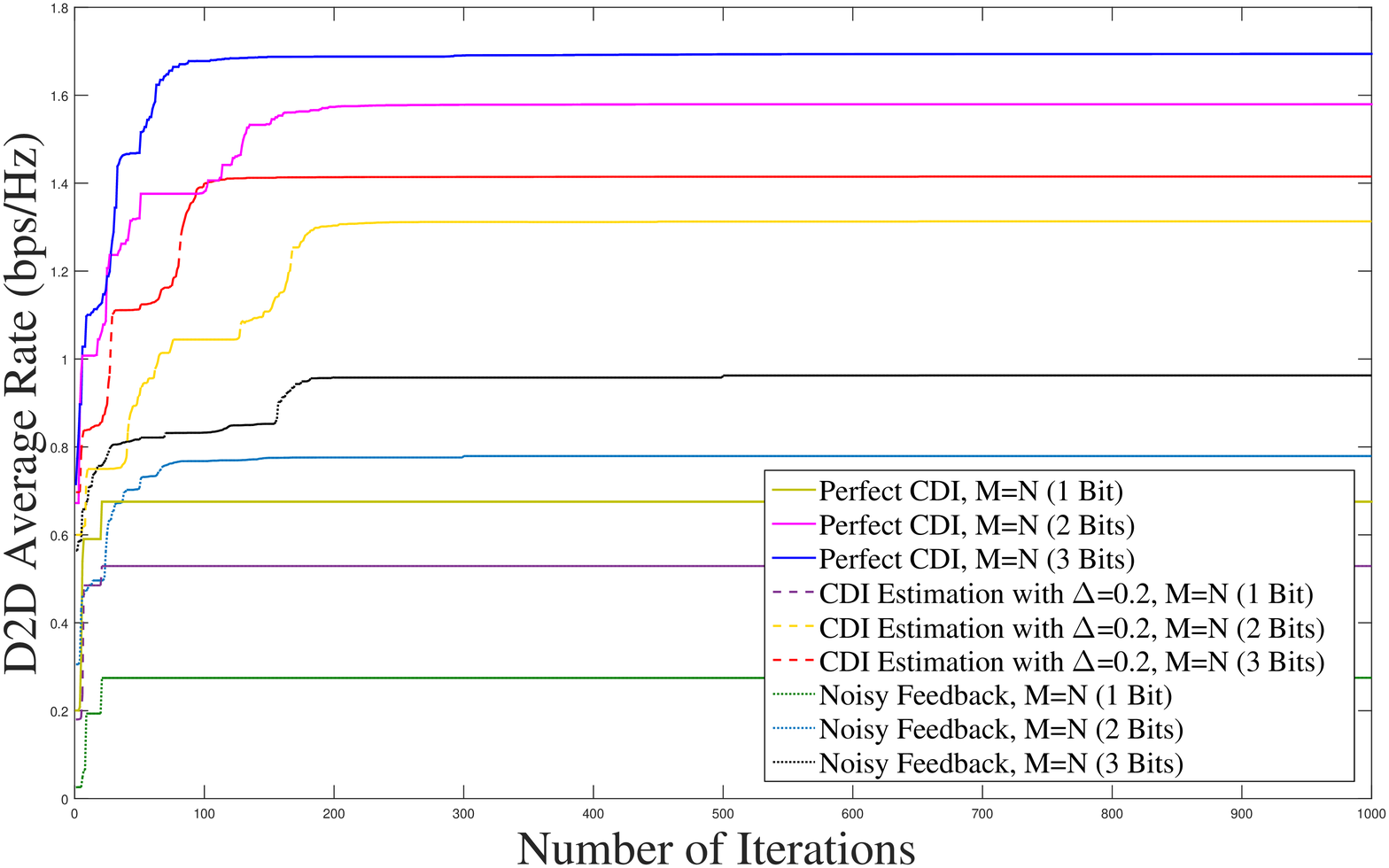} 
    \vspace{-0.3cm}
    \caption{Achieved average rate of D2D \emph{vs}. number of iteration for PSO algorithm for different feedback bits as well as error free feedback, imperfect CDI and noisy feedback.}
      \label{iter}
      \vspace{0.1cm}
    \includegraphics[scale=0.2]{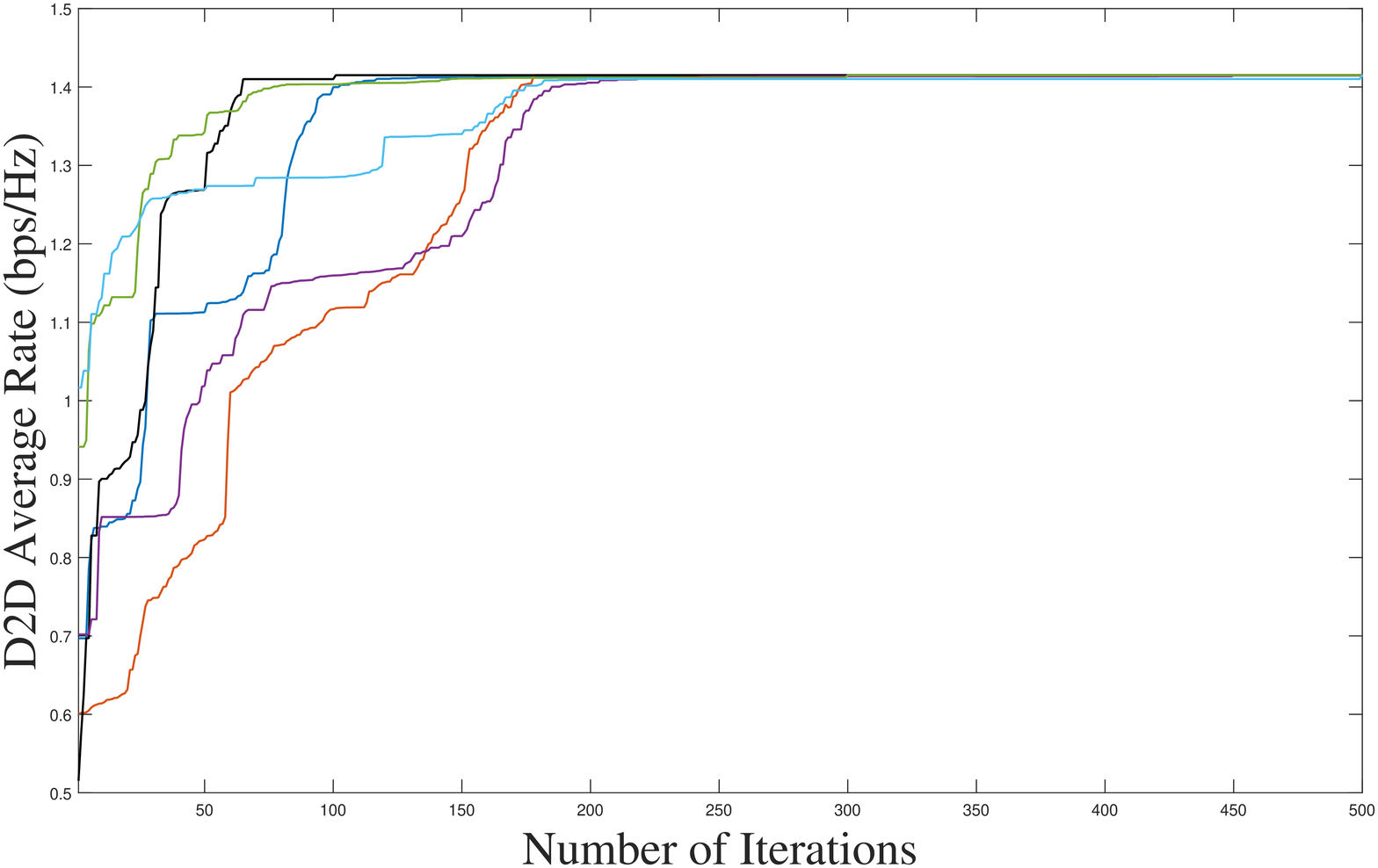} 
    \vspace{-0.3cm}
    \caption{Achieved average rate of D2D \emph{vs}. number of iteration for PSO algorithm for
    different random initialization of particles.
   System parameters: $\Delta=0.2$ and $M=N=3$ bits.}
    \vspace{-0.8cm}
      \label{iter2}
  \end{center}
\end{figure}

\vspace{-0.4cm}
\subsection{Convergence}
Fig. \ref{iter} shows the convergence of the algorithm. For the limited feedback scheme,
we consider 1, 2, and 3 bits to display the results clearly. To demonstrate the performance of the proposed
system, the results are obtained for non-noisy and noisy limited-feedback schemes for both the perfect and
imperfect CDI.
The PSO method, generally, does not guarantee to achieve global optimum for
n-dimensional functions. It is difficult to prove and show mathematically
that PSO can guarantee global optima in our problem.
However, we have used different searches to show the reliability of the PSO in
Fig. \ref{iter2}. As it is shown, with different random initialization
of swarm of particles in the different part of problem space, all of the solutions converge to the same point.

\begin{figure}[t!]
  \begin{center}
    \includegraphics[scale=0.2]{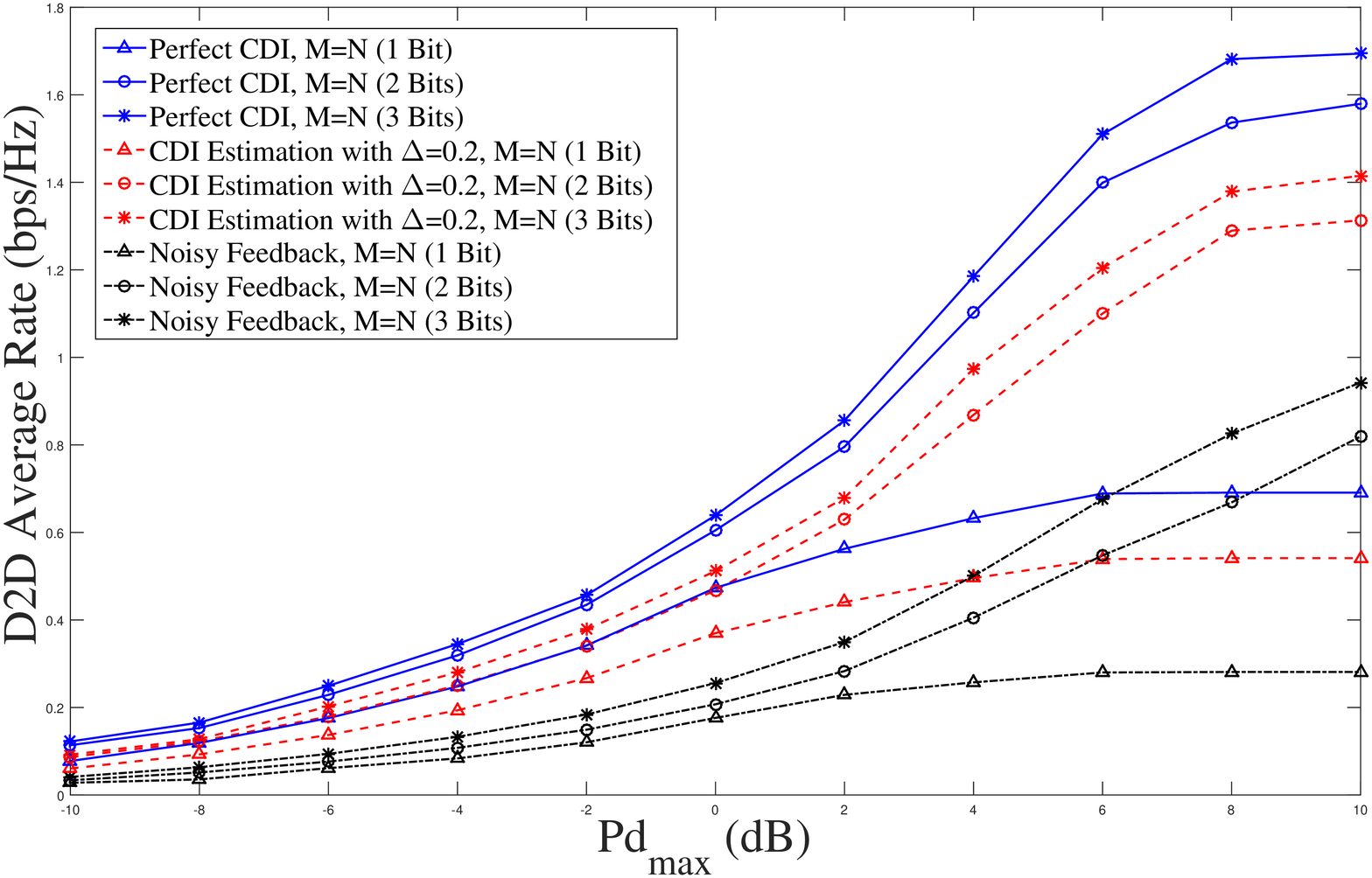} 
    \vspace{-0.3cm}
    \caption{Achieved average rate of D2D \emph{vs}. $P^{\text{D},\text{max}}$, for different feedback bits as well as error free feedback, imperfect CDI and noisy feedback.}
      \label{Pd}
      \vspace{0.4cm}
    \includegraphics[scale=0.2]{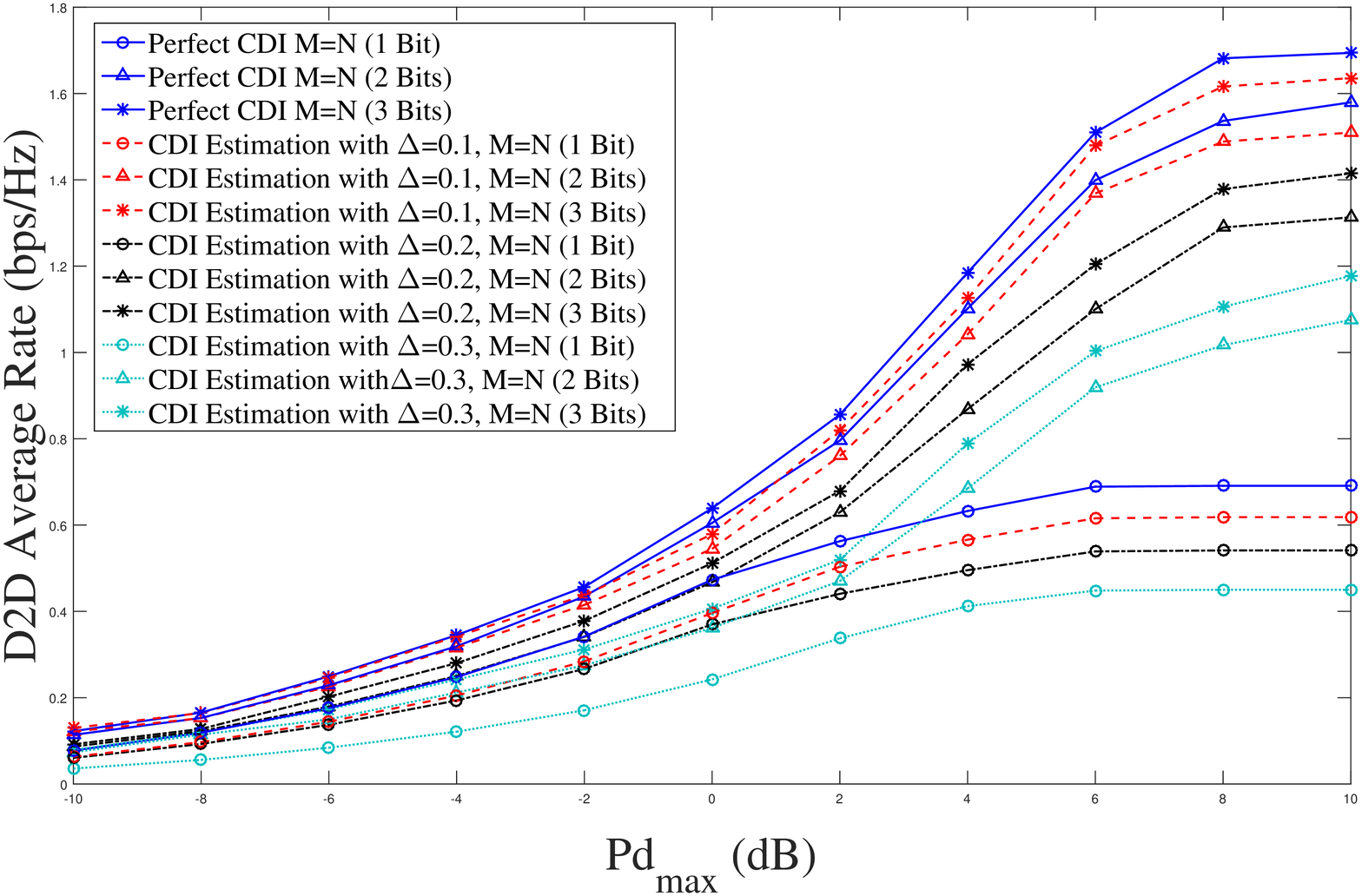} 
    \vspace{-0.3cm}
    \caption{Achieved average rate of D2D \emph{vs}. $P^{\text{D},\text{max}}$, for parametric method and different $\Delta$.}
    \vspace{-0.5cm}
      \label{param}
  \end{center}
\end{figure}

\vspace{-0.35cm}
\subsection{The Effect of the System Parameters}
In Fig. \ref{Pd} and Fig. \ref{param}, the D2D average rate is plotted versus the maximum transmit
power of D2D user ($P_\text{D}^\text{max}$) and the different number of BC and D2D feedback bits ($M, N$).
In Fig. \ref{Pd}, the D2D average rate is studied for perfect CDI and parametric CDI estimation method as well as noisy feedback. Obviously with increasing $P_\text{D}^{\text{max}}$, the
average rate of D2D increases due to increasing the feasibility set of the resource allocation problem with the relaxation of constraint on the transmit power of D2D user. As we can see, some curves are flattened when the D2D power constraint is increased. This is because the cellular rate constraint becomes the dominant factor in the optimization problem and D2D rate can not increase with increasing the transmit power. To study the effect of percent error of imperfect CDI, in Fig. \ref{param} the D2D average rate is obtained for different errors. As it is shown, by increasing the error, the D2D achievable rate decreases.

\begin{figure}[t!]
  \begin{center}
    \includegraphics[scale=0.2]{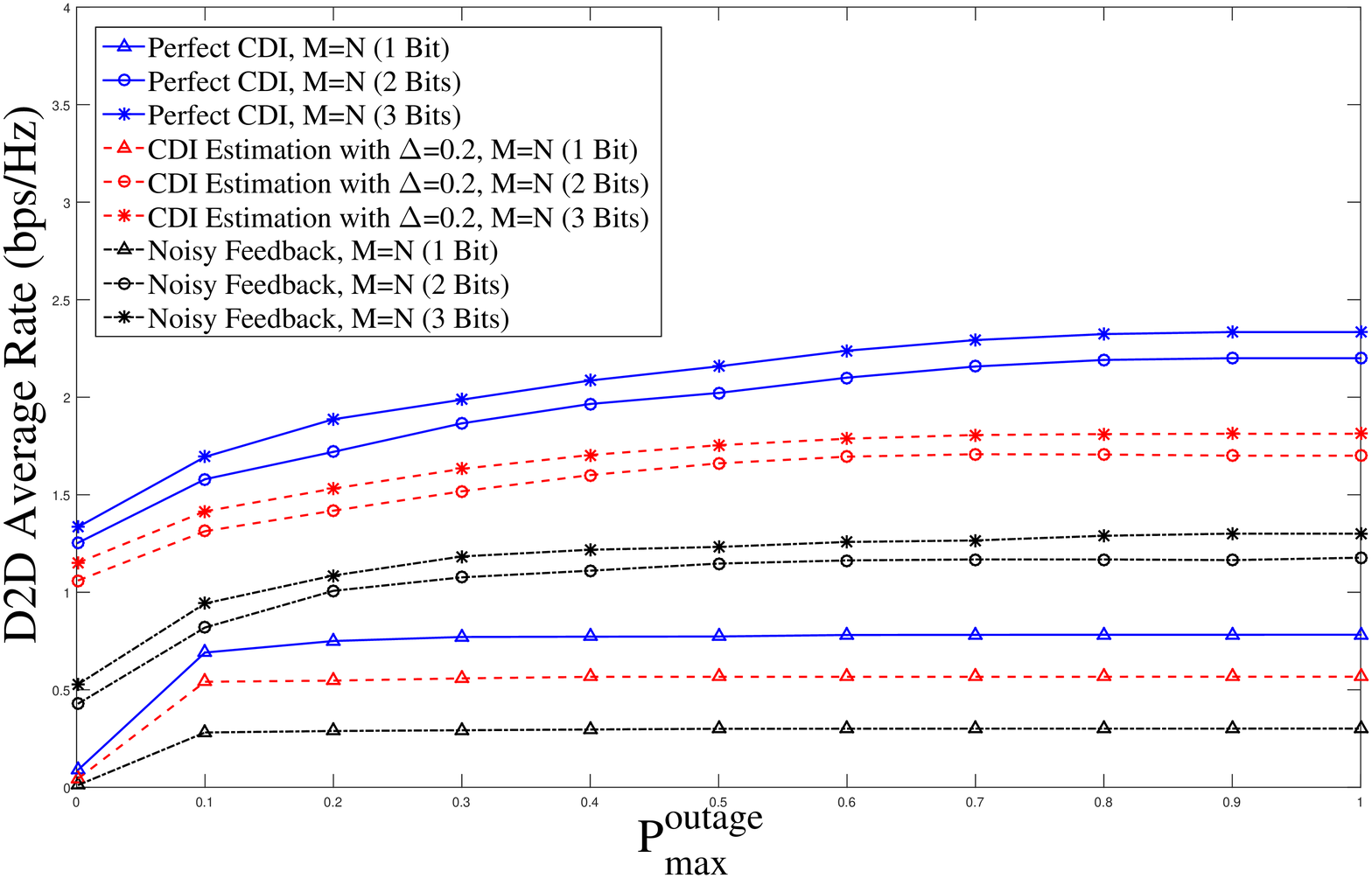} 
    \vspace{-0.4cm}
    \caption{Achieved average rate of D2D \emph{vs}.
    $P_{C^{BC}}^{\text{outage},\text{max}}$, for different
     feedback bits as well as error free feedback, imperfect CDI and noisy feedback.}
      \label{Po}
      \vspace{0.3cm}
    \includegraphics[scale=0.2]{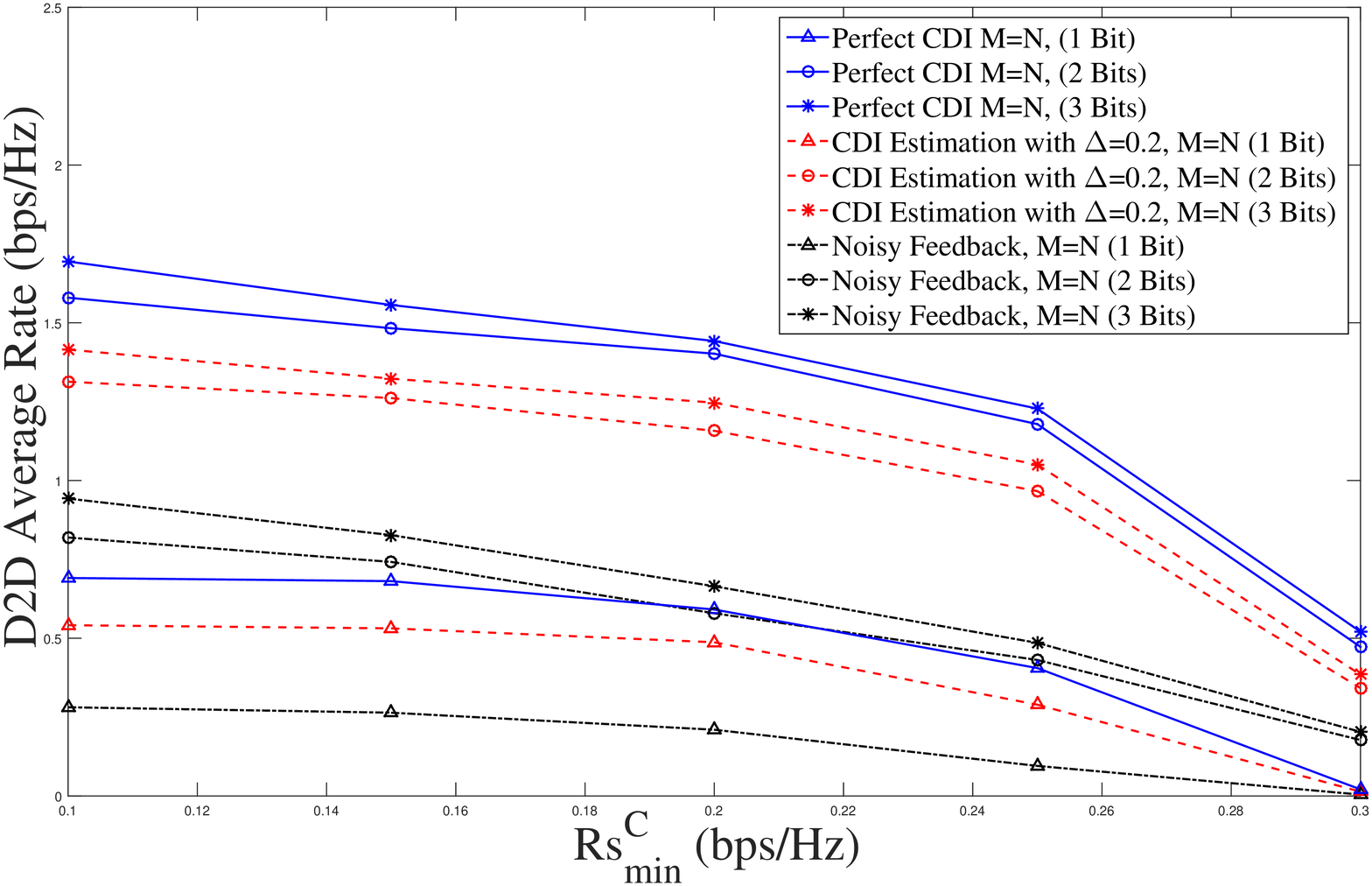} 
    \vspace{-0.3cm}
    \caption{Achieved average rate of D2D \emph{vs}. minimum secrecy rate of cellular network; $R^{\text{C}_\text{min}}_{\text{S}}$, for different feedback bits as well as error free feedback, imperfect CDI and noisy feedback.}
      \label{Rmin}
      \vspace{0.3cm}
    \includegraphics[scale=0.2]{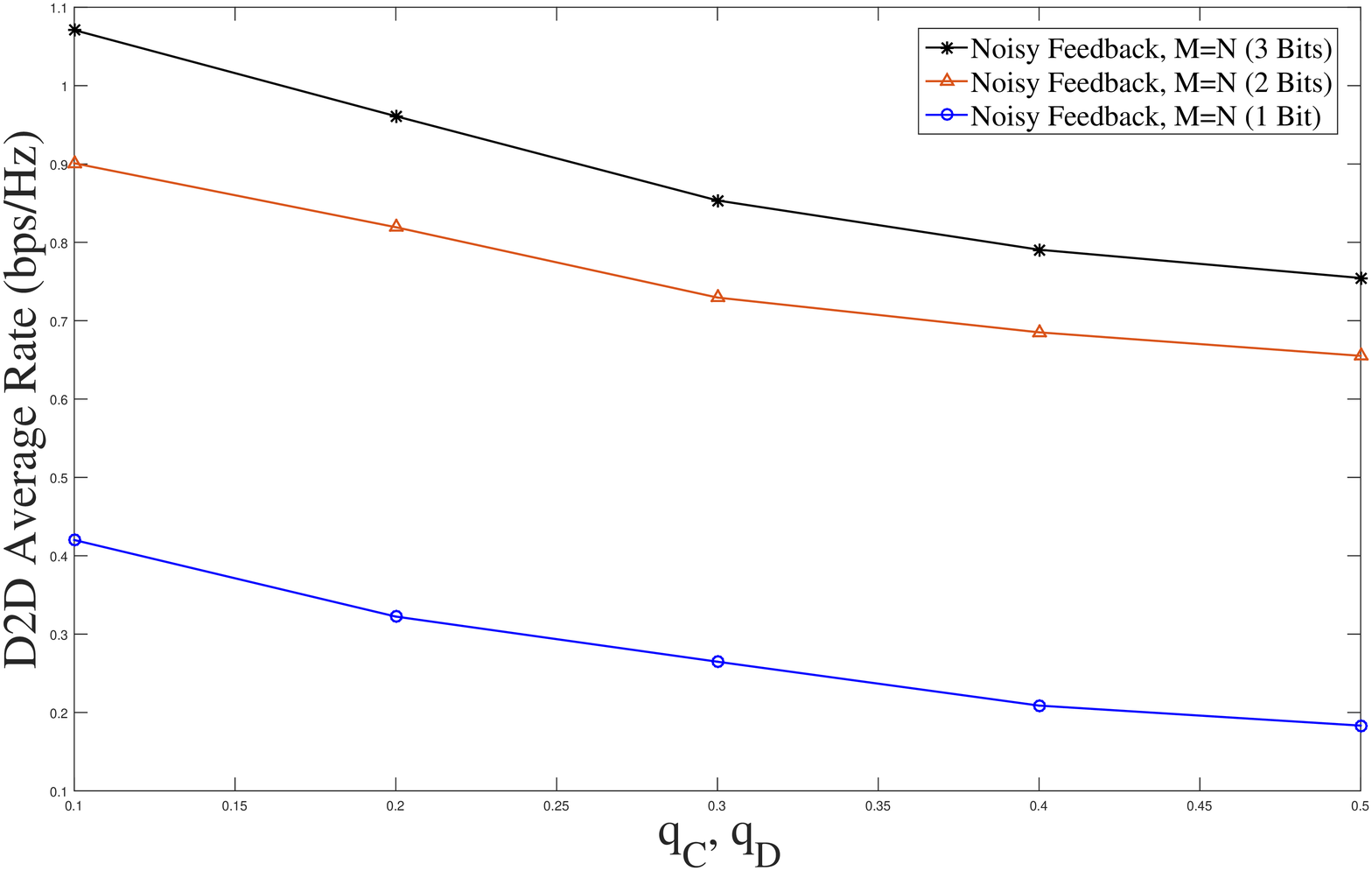} 
    \vspace{-0.3cm}
    \caption{Achieved average rate of D2D \emph{vs}. cross over probabilities of the feedback channels; $q^C, q^D$ for $M=N=2$ bits.}
      \vspace{-0.8cm}
      \label{q}
  \end{center}
\end{figure}

Fig.\ref{Po} describes the performance of D2D communication in terms of the maximum outage probability for cellular communication ($P_{\text{outage}}^{\text{max}}$) and different number of feedback bits.
 As the maximum outage probability limit increases, the D2D average rate increases. Specifically, for smaller $P_{\text{outage}}^{\text{max}}$, the overall D2D rate
increases fairly rapidly. Hence, if the cellular
communication can withstand
slight secrecy outage probability, simultaneous D2D communication
can be a great advantage. Similar to that of the previous case, the curves become flat since it is limited by D2D power constraint.

In Fig. \ref{Rmin}, the effect of the minimum required secrecy rate of the cellular network ${R_\text{S}^\text{C}}^{\text{min}}$ on the D2D rate is illustrated. Obviously, when the minimum secrecy rate of the cellular network increases, the operation of D2D communication is limited. Therefore, the D2D average rate is reduced.

In Fig. \ref{q} the effect of $q^C$ and $q^D$ is studied. As it is shown, by growing the error probability, i.e, the quality
of the feedback link degrades, we see the decline in the rate of D2D.

As it is seen in all figures, the increasing number of feedback bits results in the improvement of the D2D performance, and the average rate increases. Also, the results  demonstrate that the performances of the limited-feedback scheme without noise have the better performance in comparison with the noisy case.

\vspace{-0.35cm}
\subsection{CDI Estimation Error}
To check out the effect of CDI estimation on the performance of the system, $\frac{|\Delta r|}{r}$ is define as the percent of the difference between the average rate of D2D obtained based on the perfect and estimated CDI.

\begin{figure}[t!]
  \begin{center}
    \includegraphics[scale=0.2]{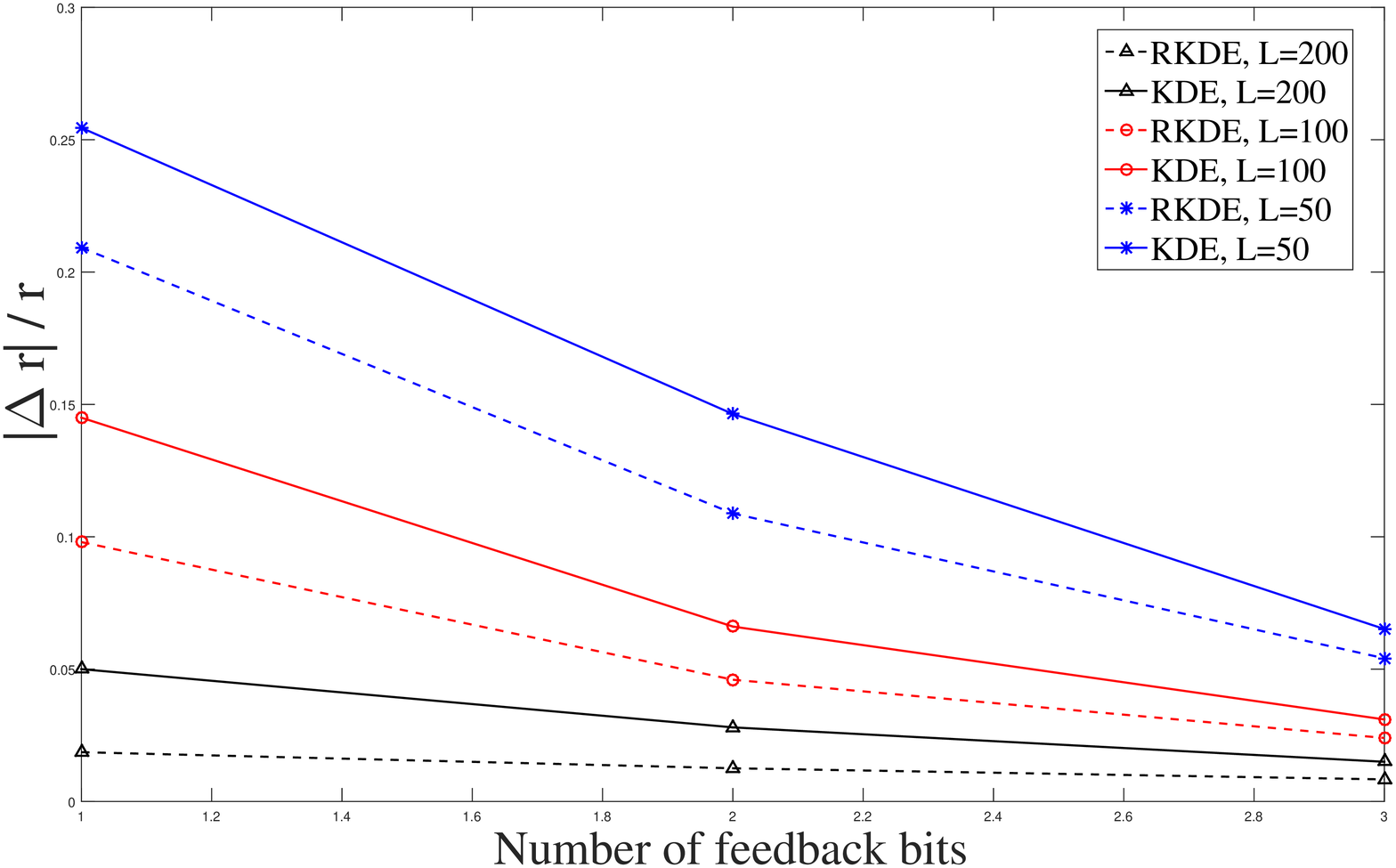} 
    \vspace{-0.4cm}
    \caption{The percent of difference between the average rate of D2D based on perfect and estimated CDI, $\frac{|\Delta r|}{r}$,  \emph{vs}. number of feedback bits , for KDE and RKDE, and different number of nominal data, $L$ and $\kappa=10$.}
      \label{nonpar_L}
    \includegraphics[scale=0.2]{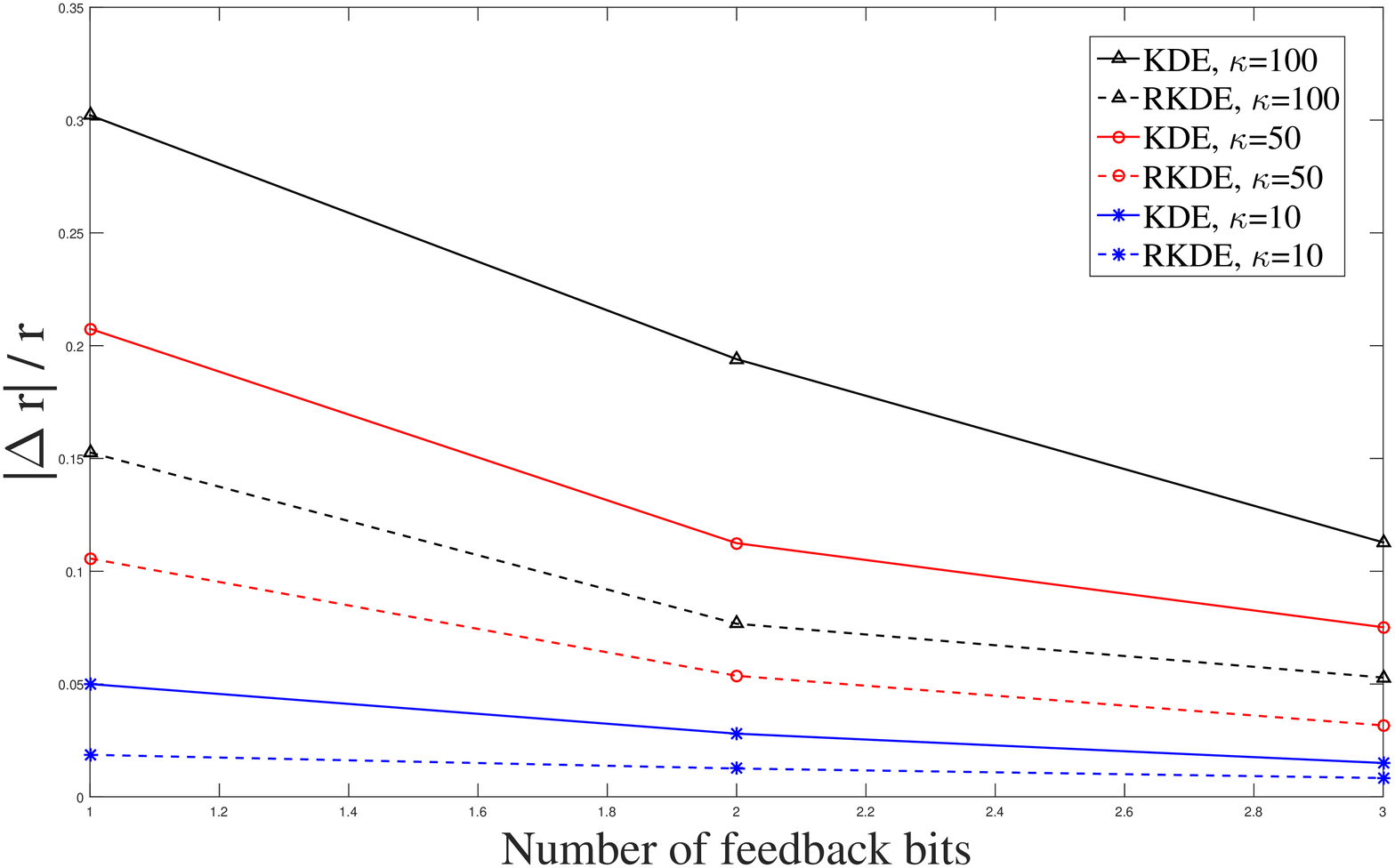} 
    \vspace{-0.3cm}
    \caption{The percent of difference between the average rate of D2D based on perfect and estimated CDI, $\frac{|\Delta r|}{r}$,  \emph{vs}. number of feedback bits , for KDE and RKDE, and different number of outlier data, $\kappa$ and $L=200$.}
     \vspace{-0.5cm}
      \label{nonpar_phi}
  \end{center}
\end{figure}

Fig.  \ref{nonpar_L} demonstrates $\frac{|\Delta r|}{r}$ as a function of the total number of users for different numbers of the nominal data where the number of outlier data is set to $\kappa=10$. As the figure shows, for small $L$ both KDE and RKDE methods perform very poor. As $L$ grows, the performance of both methods improves, and for $L = 200$, the average rate obtained based on RKDE is
very close to that of the perfect CDI case.

In Fig. \ref{nonpar_phi}, $\frac{|\Delta r|}{r}$ is plotted versus the number of feedback bits for the different number of outlier data $\kappa$ where the number of nominal data is set to $L=200$. As it is seen, the value of $\frac{|\Delta r|}{r}$ is close to zero for RKDE method with $\kappa = 10$. As $\kappa$ grows, $\frac{|\Delta r|}{r}$ increases implying the divergence from the actual pdf. It is also observed that the value
of $\frac{|\Delta r|}{r}$ for KDE is far away from zero and the performance degrades faster compared to that of RKDE
as $\kappa$ grows.

\vspace{-0.3cm}
\section{Conclusions}\label{conclusions}
%
In this paper, we studied a limited-feedback radio resource
allocation problem for the D2D communication scenario underlaying
an existing cellular network with the objective of maximizing the
D2D average rate subject to average users transmit power limitations, the average secrecy rate and outage
probability threshold for the cellular network. Through the PSO algorithm, the appropriate
code book for the channel partitioning was designed. In addition, we
solved the problem when the feedback
channel is noisy. To investigate the effect of the CDI imperfectness on the
performance, we applied both the parametric and non-parametric methods.
Using simulations, we studied the impact of the system
parameters, such as the  maximum allowable transmit power of D2D
user, the number of feedback bits, and the minimum secrecy rate of
cellular network, on the achievable rate of D2D. As it was shown, by
more feedback bits, better D2D performance can be achieved.

\vspace{-0.3cm}
\appendices \numberwithin{equation}{section}
\setcounter{equation}{0}

\section{Finding Outage Probability in (\ref{outageprobabilitycellularregionmn})}\label{findingprobabilityoutageprobabilitycellularregionmn}
To compute the outage probability in \eqref{outageprobabilitycellularregionmn}, we should compute the success probability in \eqref{successprobabilitycellularregionmn}. Note that, we have $h^{\text{BC}}\in \mathcal{R}^{\text{BC}}_m=
[\widetilde{h}^{\text{BC}}(m),\widetilde{h}^{\text{BC}}(m+1)]$ and $h^{\text{DD}}\in \mathcal{R}^{\text{DD}}_n=
[\widetilde{h}^{\text{DD}}(n),\widetilde{h}^{\text{DD}}(n+1)]$. The success probability can be written as follows

\vspace{-0.3cm}
\begin{align}\nonumber
&P^\text{success}_{p^{\text{BC}}(m),
r^{\text{BC}}(m),r_\text{S}^{\text{BC}}(m),p^{\text{DD}}(n),
r^{\text{DD}}(n)} \\\nonumber&=\Pr\bigg(r^{\text{BC}}(m)\leq
\hat{C}^{\text{C}}(m,n),\hat{C}^\text{BE}(m,n)\leq r^{\text{e}}(m)\bigg)\\\nonumber&
=\Pr\bigg(\widetilde{h}^{\text{BC}}(m)\leq
\hat{h}^{\text{BC}},  \hat{h}^{\text{BE}}\leq 2^{r^{\text{e}}(m)}-1
 \bigg)\\\nonumber&=\Pr\bigg(\widetilde{h}^{\text{BC}}(m)\leq
\hat{h}^{\text{BC}}\bigg) \Pr\bigg(\hat{h}^{\text{BE}}\leq 2^{r^{\text{e}}(m)}-1
 \bigg)\\\label{successprobability1}&=\Big(1-F^{m,n}_{\hat{h}^{\text{BC}}}(\widetilde{h}^{\text{BC}}(m))\Big){F}_{\hat{h}^{\text{BE}}}^n(2^{r^{\text{e}}(m)}-1),
\end{align}
where $ \hat{h}^{\text{BC}} =\frac{h^{\text{BC}}}{1+h^{\text{DC}}p^{\text{DD}}(n)}$ depends on $m$ and $n$, $\hat{h}^{\text{BE}}=\frac{h^{\text{BE}}}{1+h^{\text{DE}}p^{\text{DD}}(n)}$ depends on $n$, and the variables $\hat{h}^{\text{BC}}$ and $\hat{h}^{\text{BC}}$ are independent and hence the product term in \eqref{successprobability1} follows. Note that in the above equations, we have implicitly assumed that $h^{\text{BC}}\in \mathcal{R}^{\text{BC}}_m$ and $h^{\text{DD}}\in \mathcal{R}^{\text{DD}}_n$ meaning that the above probability is a conditional probability.

%
%

To compute \eqref{successprobability1}, we need to find PDFs of $\hat{h}^{\text{BC}}$ and $\hat{h}^{\text{BE}}$. The CDF of $\hat{h}^{\text{BC}}$ can be computed as follows:

\vspace{-0.3cm}
\begin{align}\nonumber
&F^{m,n}_{\hat{h}^{\text{BC}}}(x)=\Pr\big(\hat{h}^{\text{BC}}\leq
x| h^{\text{BC}}\!\in\!\mathcal{R}^{\text{BC}}_m,h^{\text{DD}}\!\in\! \mathcal{R}^{\text{DD}}_n\big)\\\nonumber &=\!\Pr\big(\frac{h^{\text{BC}}}{1+h^{\text{DC}}p^{\text{DD}}(n)}\leq
x | h^{\text{BC}}\!\in\!\mathcal{R}^{\text{BC}}_m,h^{\text{DD}}\!\in\! \mathcal{R}^{\text{DD}}_n\big)\\\nonumber
&=\!
\Pr\big(h^{\text{BC}}\leq(1+h^{\text{DC}}p^{\text{DD}}(n))x | h^{\text{BC}}\!\in\! \mathcal{R}^{\text{BC}}_m,h^{\text{DD}}\in \mathcal{R}^{\text{DD}}_n\big)\\
\nonumber
&=\!\left\{ \begin{array}{ll}
\!\!\Pr\bigg(h^{\text{BC}}\!\leq\! x | h^{\text{BC}}\!\in\! \mathcal{R}^{\text{BC}}_m,h^{\text{DD}}\!\in\!\mathcal{R}^{\text{DD}}_n\bigg) & \!\!\!\mbox{\text{if} $n= 0$},\\
\!\!\Pr\!\bigg(\frac{\frac{h^{\text{BC}}}{x}-1}{p^{\text{DD}}(n)}\!\leq\! h^{\text{DC}} | h^{\text{BC}}\!\in\!\mathcal{R}^{\text{BC}}_m,h^{\text{DD}}\!\in\! \mathcal{R}^{\text{DD}}_n\bigg) & \!\!\mbox{\text{if} $n \neq 0$}.\end{array} \right.\\\label{cdfhathbcmn12}
&
\end{align}

\vspace{-0.3cm}
In \eqref{cdfhathbcmn12}, for the cases $n=0$ and $n\neq 0$, respectively, we have the followings:

\vspace{-0.3cm}
\begin{align} \nonumber
&F^{m,0}_{\hat{h}^{\text{BC}}}(x)=\frac{\Pr\bigg(h^{\text{BC}}\leq
x,h^{\text{BC}}\in \mathcal{R}^{\text{BC}}_m
\bigg)}{\Pr\bigg(h^{\text{BC}}\in \mathcal{R}^{\text{BC}}_m
\bigg)}=
 \frac{1}{\Pr\bigg(h^{\text{BC}}\in \mathcal{R}^{\text{BC}}_m\bigg)} \\\label{cdfhathbcmn21}
&\times \!\bigintsss_{\widetilde{h}^{\text{BC}}(m)}^{x} \!\!\!\!\!f_{h^{\text{BC}}}(h^{\text{BC}}) dh^{\text{BC}},\:\mbox{\text{if} $\widetilde{h}^{\text{BC}}(m) \!\leq x \!\leq \widetilde{h}^{\text{BC}}(m+1)$},
 \end{align}

\vspace{-0.5cm}
\begin{align}\nonumber
  & F^{m,n}_{\hat{h}^{\text{BC}}}(x)=\\\nonumber
  &\frac{\Pr\bigg(\frac{\frac{h^{\text{BC}}}{x}-1}{p^{\text{DD}}(n)}\leq h^{\text{DC}}, h^{\text{BC}}\in \mathcal{R}^{\text{BC}}_m\bigg)}{\Pr\bigg(h^{\text{BC}}\in \mathcal{R}^{\text{BC}}_m\bigg)}=
 \frac{1}{\Pr\bigg(h^{\text{BC}}\in \mathcal{R}^{\text{BC}}_m\bigg)}\\ \nonumber &\times\!\bigintsss_{\widetilde{h}^{\text{BC}}(m)}^{\widetilde{h}^{\text{BC}}(m+1)} \!\!\bigintsss_{\frac{\frac{h^{\text{BC}}}{x}-1}{p^{\text{DD}}(n)}}^{\infty}  f_{h^{\text{DC}}}(h^{\text{DC}}) f_{h^{\text{BC}}}(h^{\text{BC}}) dh^{\text{DC}} dh^{\text{BC}},\\\label{cdfhathbcmn22}
& \qquad\qquad\qquad \mbox{\text{if} $0< x \leq \widetilde{h}^{\text{BC}}(m+1)$}.
\end{align}
The CDF of $\hat{h}^{\text{BE}}$ is given by
\begin{align}\nonumber
&{F}_{\hat{h}^{\text{BE}}}^n(x) = 1-\frac{\bar{h}^{\text{BE}}}{\bar{h}^{\text{BE}}+\bar{h}^{\text{DE}}p^{\text{DD}}(n)x}
  \exp \bigg( -\frac{x}{\bar{h}^{\text{BE}}}\bigg), \\\label{cdfhathbemn2}
  &\qquad\qquad\qquad\qquad\qquad\qquad \mbox{\text{if} $0\leq x <\infty $}.
\end{align}
Finally, we will have the following:
 \begin{align}
 \nonumber
 & P^\text{outage}_{\mathcal{C}^\text{BC}} = \sum_{n=1}^{N-1} \sum_{m=1}^{M-1}
   \Pr\bigg(h^{\text{DD}}\!\in\!
[\widetilde{h}^{\text{DD}}(n),\widetilde{h}^{\text{DD}}(n+1))\bigg)\\ \nonumber
&\times
 \Pr\bigg(h^{\text{BC}}\!\in\!
[h^{\text{BC}}(m),h^{\text{BC}}(m+1))\bigg)\\
&\times P^{success}_{p^{\text{BC}}(m), r^{\text{BC}}(m),r_\text{S}^{\text{BC}}(m),p^{\text{DD}}(n), r^{\text{DD}}(n)}.
 \end{align}

\section{Finding Success Probability in (\ref{averageratecellulard2dabsent})} \label{findingsuccessprobabilityinaverageratecellulard2dabsent}
To obtain the success probability in
(\ref{averageratecellulard2dabsent}), i.e.,
$\Pr\bigg(h^{\text{BC}}\in \mathcal{R}^{\text{BC}}_m,r_\text{S}^{\text{BC}}(m)\leq
C_\text{S}^{\text{C}}(m)\bigg)$, we first define new random
variables $x_m=1+h^{\text{BC}} p^{\text{BC}}(m)$ and
$y_m=1+h^{\text{BE}} p^{\text{BC}}(m)$ with respective
distributions
$f_{x_m}(x_m)=\frac{1}{p^{\text{BC}}(m)}f_{h^{\text{BC}}}\bigg(\frac{x_m-1}{p^{\text{BC}}(m)}\bigg)$
and
$f_{y_m}(y_m)=\frac{1}{p^{\text{BC}}(m)}f_{h^{\text{BE}}}\bigg(\frac{y_m-1}{p^{\text{BC}}(m)}\bigg)$.
Therefore, we have
\begin{align}\nonumber
&\Pr\bigg(r_\text{S}^{\text{BC}}(m)\leq
C_\text{S}^{\text{C}}(m),h^{\text{BC}}\in \mathcal{R}^{\text{BC}}_m)\bigg)\\ \nonumber
&=Pr\bigg(y_m\leq
2^{-r_\text{S}^{\text{BC}}(m)} x_m  ,h^{\text{BC}}\in \mathcal{R}^{\text{BC}}_m\bigg)\\\label{secrecysuccessprobabilityregioncellularD2Dabsent1}&
=\int_{1+\widetilde{h}^{\text{BC}}(m)
p^{\text{BC}}(m)}^{1+\widetilde{h}^{\text{BC}}(m+1)
p^{\text{BC}}(m)}\int_{1}^{2^{-r_\text{S}^{\text{BC}}(m)}
x_m}\!\!\!f_{y_m}(y_m) f_{x_m}(x_m) dy_m dx_m.
\end{align}

\section{Finding Success Probability in (\ref{averagerateD2Dcellularpresentconficetial})}\label{findingprobabilityaveragerateD2Dcellularpresentconficetial}
To obtain the success probability in (\ref{averagerateD2Dcellularpresentconficetial}), i.e.,
$ \Pr\bigg(h^{\text{BC}}\in \mathcal{R}^{\text{BC}}_m,h^{\text{DD}}\in \mathcal{R}^{\text{DD}}_n,r^{\text{DD}}(n)\leq
\hat{C}^{\text{D}}(n)\bigg)$,
first, we introduce the cumulative distribution function of
 $h$, which is equal to $F^{h}_{m}(x)=1-e^{\frac{-x}{\bar{h}}}$ and
 \begin{align}
&G^{h}_{m}\!= F^{h}_{m}\!\bigg(\!\widetilde{h}(m+1)\!\bigg)\!-\!F^{h}_{m}\!\bigg(\!\widetilde{h}(m)\!\bigg)\!
\!=\!e^{\frac{-\widetilde{h}(m)}{\bar{h}}}\!-\!e^{\frac{-\widetilde{h}(m+1)}{\bar{h}}},
\end{align}
which is the probability that $h$ falls into the region of $[\widetilde{h}(m),\widetilde{h}(m+1))$. Therefre, for $h^{\text{BC}}$ we have the following:
\begin{align}\nonumber
&G^{h^{\text{BC}}}_{m}=F^{h^{\text{BC}}}_{m}\bigg(\widetilde{h}^{\text{BC}}(m+1)\bigg)-F^{h^{\text{BC}}}_{m}\bigg(\widetilde{h}^{\text{BC}}(m)\bigg )\\
&=e^{\frac{-\widetilde{h}^{\text{BC}}(m)}{\bar{h}^{\text{BC}}}}-e^{\frac{-\widetilde{h}^{\text{BC}}(m+1)}{\bar{h}^{\text{BC}}}}.
\end{align}
Then, we have the following:
\begin{align}\nonumber
  &\Pr\bigg(h^{\text{BC}}\in \mathcal{R}^{\text{BC}}_m,h^{\text{DD}}\in \mathcal{R}^{\text{DD}}_n,r^{\text{DD}}(n)\leq
\hat{C}^{\text{D}}(n)\bigg)\\
&
 =G^{h^{\text{BC}}}_{m} \Pr\bigg(h^{\text{DD}}\in\mathcal{R}^{\text{DD}}_n,\widetilde{h}^{\text{DD}}(n)\leq
\hat{h}^{\text{DD}}\bigg),
\end{align}
and given that $\hat{h}^{\text{DD}}=\frac{h^{\text{DD}}}{1+h^{\text{BD}}p^{\text{BC}}(m)}$,
we will have the following:
\begin{align}\nonumber
 & \Pr\bigg(\widetilde{h}^{\text{DD}}(n)\leq
\hat{h}^{\text{DD}}, h^{\text{DD}}\in \mathcal{R}^{\text{DD}}_n\bigg)
  \\\nonumber
  &=\Pr\bigg(0\leq h^{\text{BD}}\leq
\frac{(\frac{h^{\text{DD}}}{\widetilde{h}^{\text{DD}}(n)}-1)}{p^{\text{BC}}(m)}, h^{\text{DD}}\in \mathcal{R}^{\text{DD}}_n\bigg) \\&
=\int^{\widetilde{h}^{\text{DD}}(n+1)}_{\widetilde{h}^{\text{DD}}(n)} \bigg[1-e^{\frac{\bigg(\frac{h^{\text{DD}}}{\widetilde{h}^{\text{DD}}(n)}-1\bigg)}{\bar{h}^{\text{BD}}p^{\text{BC}}(m)}} \bigg] \frac{1}{\bar{h}^{\text{DD}}} e^{-\frac{h^{\text{DD}}}{\bar{h}^{\text{DD}}}} d {h^{\text{DD}}}.
\end{align}
Finally, the success probability in (\ref{averagerateD2Dcellularpresentconficetial}) is equal to:
\begin{align}\nonumber
&\Pr\bigg(h^{\text{BC}}\in
\mathcal{R}^{\text{BC}}_m,h^{\text{DD}}\in
\mathcal{R}^{\text{DD}}_n,r^{\text{DD}}(n)\leq
\hat{C}^{\text{D}}(n)\bigg)\\
&\!=\!G^{h^{\text{BC}}}_{m}
\!\int^{\widetilde{h}^{\text{DD}}(n+1)}_{\widetilde{h}^{\text{DD}}(n)}
\!\bigg[1-e^{\frac{\bigg(\frac{h^{\text{DD}}}{\widetilde{h}^{\text{DD}}(n)}-1\bigg)}{\bar{h}^{\text{BD}}p^{\text{BC}}(m)}}
\bigg] \frac{1}{\bar{h}^{\text{DD}}}
e^{-\frac{h^{\text{DD}}}{\bar{h}^{\text{DD}}}}d{h^{\text{DD}}}.
\end{align}

\bibliographystyle{IEEEtran}
\bibliography{IEEEabrv,Bibliography}

\end{document}